\documentclass[a4paper,11pt,feynmf]{article}
\usepackage{graphicx}
\usepackage{amsmath,amssymb}
\usepackage{graphics,color,array,calc,rotating,epsfig,psfrag}
\numberwithin{equation}{section}
\usepackage{cite}
\usepackage{bm}
\usepackage{dcolumn}
\usepackage{float}
\usepackage{subfigure}
\usepackage{amsfonts}
\usepackage[mathscr]{eucal}
\def\be{\begin{equation}} \def\ee{\end{equation}}
\def\bea{\begin{eqnarray}} \def\eea{\end{eqnarray}}

\begin{document}
	%
	\baselineskip 18pt%
	\begin{titlepage}
		\vspace*{1mm}%
		\hfill%
		\vspace*{15mm}%
		\hfill
		\vbox{
			\halign{#\hfil         \cr
			} 
		}  
		\vspace*{20mm}
		
		\begin{center}
			{\large {\bf The utilization of magneto-hydrodynamics framework in
					expansion of   magnetized viscous conformal flow
			}}\\
			\vspace*{5mm}
			{  M.~Karimabadi\footnote{ma.karimabadi@hsu.ac.ir}, A.F.~Kord\footnote{a.f.kord@hsu.ac.ir}, B.~Azadegan\footnote{azadegan@hsu.ac.ir
			} }\\
			
			\vspace*{0.2cm}
			{$^{}$ Department of Physics, Hakim Sabzevari University, P.O. Box 397, Sabzevar, Iran}\\
			\vspace*{1cm}
		\end{center}
		\begin{abstract}
			
			The evolution of magnetized quark gluon plasma (QGP) in the framework of magneto-hydrodynamics is the focus of our study. We are investigating the temporal and spatial evolution of QGP using a second order viscous hydrodynamic framework. The fluid is considered to be magnetized and subjected to the influence of a magnetic field that is generated during the early stages of relativistic heavy ion collisions.We assume boost invariance along the beam line, which is represented by the $z$ coordinate, and fluid expansion in the $x$ direction. Additionally, we assume that the magnetic field is perpendicular to the reaction plane, which corresponds to the $y$ direction. The fluid is considered to have infinite electrical conductivity. To analyze this system, we solve the coupled Maxwell and conservation equations. By doing so, we are able to determine the time and space dependence of the energy density, velocity, and magnetic field in the transverse plane of the viscous magnetized hot plasma. Furthermore, we obtain the spectrum of hadrons and compare it with experimental data.
			\end{abstract}
		
	\end{titlepage}
	\section{introduction}
Nowadays, it is widely acknowledged that when relativistic heavy-ions collide, they produce a highly energetic and dense matter known as a fireball. Within this fireball, quarks and gluons exist in a deconfined state referred to as a quark-gluon plasma (QGP). This state is achieved after the initial collisions of the hard partons from the nuclei.

The evolution of matter generated in ultra-relativistic collisions is governed by fluid dynamics. Consequently, relativistic hydrodynamics models that incorporate finite viscous correction play a crucial role in comprehending the space-time evolution of the system formed during such collisions.
 The hydrodynamic approach has proven to be one of the most effective methods for describing the QGP matter. In particular, it has provided reliable estimates for the lowest shear viscosity to entropy ratio. This theoretical framework has demonstrated consistent results when compared to numerous experimental findings~\cite{na1}-\cite{na6}.

 During the last decade,  the principles of relativistic viscous fluid dynamics have been developed within the framework of an effective field theory. This formulation is solely based on the understanding of symmetries and long-lasting degrees of freedom. Coincidentally, during the same period, numerical simulations for the matter produced in relativistic heavy-ion collision experiments were initially made accessible. These simulations have played a crucial role in establishing limitations on the shear viscosity in Quantum Chromodynamics (QCD).
Besides, The AdS/CFT correspondence has provided a prediction for the lower limit of the shear viscosity $(\eta)$ to entropy density $(s)$ ratio, which is $\eta/s = 1/4\pi$ \cite{na7}. Notably, hydrodynamic models with $\eta/s = 0.2$ effectively account for the observed elliptic flow results obtained from the RHIC experiments\cite{na7}.

Furthermore, when relativistic heavy ions collide, not only does it lead to the formation of hot dense nuclear matter, but it also generates an immensely powerful magnetic field on the order of $eB \ 10^{18} - 10^{19}\ G$. This magnetic field is approximately $10^{13}$ times stronger than any magnetic field that has ever been produced in a laboratory setting. The interaction between this magnetic field and the Quark-Gluon Plasma (QGP) is predicted to give rise to various intriguing phenomena, such as the Chiral Magnetic Effect (CME), Chiral Magnetic wave (CMW), Chiral Electric Separation Effect (CESE), Chiral Hall Separation Effect (CHSE), pressure anisotropy in the QGP, influence on the direct and elliptic flow, and a shift in the critical temperature. For further information and in-depth analysis on these subjects, please refer to the relevant references and review articles listed in Refs\cite{a13,a14,a15,a16,a17,a18,a19,a20,a21}.

In order to study the development of QGP matter, it is customary to utilize relativistic hydrodynamics, a well-established and effective approach~ \cite{na1,na2,a22,a23,a24}. Furthermore, to investigate the influence of magnetic fields on QGP matter, it is crucial to establish and employ a framework for relativistic magneto-hydrodynamics (RMHD). Indeed, the RMHD framework serves as a vital tool for characterizing hot plasma in the presence of electromagnetic fields \cite{a25,a26}. Consequently, a numerical code capable of solving the equations of (1+3) dimensional relativistic magneto-hydrodynamics becomes indispensable for this purpose.

 The primary objective of this investigation is to examine and analyze the reaction of a viscous fluid that possesses a non-zero magnetization when subjected to a coupled magnetic field. Consistent with previous studies, we make the assumption that a magnetic field is generated in the $y$ direction, which is perpendicular to the reaction plane. Additionally, we assume that a four-velocity profile exhibits a non-zero component in the $x$ direction. Furthermore, we maintain the assumption of boost invariance along the beam line. In order to accomplish our research goal, we solve a system of coupled equations known as the resistive magnetohydrodynamic (RMHD) equations. The purpose of this study is to investigate the potential influence of matter viscosity on the magnetohydrodynamic (MHD) evolution of the quark-gluon plasma (QGP). Moreover, we aim to explore the effects of the magnetic field on the pressure anisotropy of the fluid. Additionally, we will present a comparison between the results of our calculation for the hadron spectrum and the experimental data.

\section{Conformal Second Order Hydrodynamics}
In this section, we provide a concise overview of our formalism for describing the behavior of viscous  fluid in the presence of an uncharged medium with nonzero magnetization. Additionally, we take into account the interaction between the QGP medium and electromagnetic fields. To simplify the analysis, we assume a dissipative relativistic plasma with infinite electrical conductivity, resulting in the disappearance of the electric field in the local rest-frame. Furthermore, we adopt a linear approximation to describe the magnetization effect, which can be characterized by the magnetic susceptibility $\chi_m$. Under this linear approximation, the magnetic polarization can be expressed.

\begin{eqnarray}
\label{1}
 M^{\mu}&=&\chi_m B^{\mu}.
\end{eqnarray}
For a fluid with
infinite electrical conductivity,  the covariant form of  of RMHD equations  are given by:

\begin{eqnarray}\label{2}
\nabla_\mu(T_{(0)F}^{\mu\nu}+T_{EM}^{\mu\nu}+\pi^{\mu\nu}+ \Delta^{\mu \nu} \Pi )&=&0,\\ \label{3}
\nabla_\mu F^{*\mu\nu}&=&0.
\end{eqnarray}
The quantities $\pi^{\mu\nu}$ and $\Pi$ are commonly known as the shear stress and bulk stress, respectively. These terms represent corrections in the context of viscous hydrodynamics, where $\pi^{\mu\nu}$ corresponds to the traceless part and $\Pi$ corresponds to the trace part. In this study, we examine a viscous fluid that exhibits conformal properties. Consequently, it is necessary for the trace component of the energy-momentum tensor to be zero ($\Pi=0$). A specific category of theories exists in which the expectation value of the trace of the energy-momentum tensor vanishes in Minkowski space. These theories are commonly known as conformal theories. The energy momentum tensor for the viscous conformal magnetism fluid in a flat space can be expressed as follows\cite{new1}:

\begin{eqnarray}\label{4}
T_{F}^{\mu\nu}&=& T_{(0)F}^{\mu\nu}+\pi
^{\mu \nu},\\ \label{5}
T_{(0)F}^{\mu\nu}&=&(\epsilon+P-M)u^\mu u^\nu+(P-M)g^{\mu\nu}+M^{\mu}B^{\nu},\\ \label{6}
\pi
^{\mu \nu}&= & -\eta \sigma^{\mu \nu}+\eta \tau_\pi\left[{ }^{<} D \sigma^{\mu \nu>}+\frac{\nabla \cdot u}{3} \sigma^{\mu \nu}\right]+\lambda_1 \sigma^{<\mu}{ }_\lambda \sigma^{\nu>\lambda}+\lambda_2 \sigma_\lambda^{<\mu} \Omega^{\nu>\lambda}\\&& \nonumber +\lambda_3 \Omega^{<\mu} \Omega^{\nu>\lambda}, 
\end{eqnarray}
here, 

\begin{eqnarray}\label{7}
\sigma^{\mu \nu}&=&2 \nabla^{<\mu} u^{\nu>}=2 \nabla_{\perp}^{(\mu} u^{\nu)}-\frac{2}{3} \Delta^{\mu \nu} \nabla_\lambda^{\perp} u^\lambda, \\ \label{8}
\Omega_{\mu\nu}&=&\nabla_{[\mu}^{\perp} u_{\nu]}=\frac{1}{2}\left(\nabla_\mu^{\perp} u_\nu-\nabla_\nu^{\perp} u_\mu\right).
\end{eqnarray}
The expression $ \nabla^{(\mu} u^{\nu)}$ is defined using the notation $A^{(\mu} B^{\nu)}=\frac{1}{2}(A^{\mu} B^{\nu}+A^{\nu} B^{\mu})$. Additionally, we have $D=u^{\mu} \nabla_{\mu}$, $\nabla_{\perp}^{\rho}=\Delta^{\mu \rho} \nabla_{\mu}$, and $\Delta^{\mu \nu}= g^{\mu \nu}+ u^{\mu} u^{\nu}$.
Besides, the electromagnetic tensor is given by:
\begin{eqnarray}\label{9}
T_{EM}^{\mu\nu}&=&B^2 u^\mu u^\nu+\frac{1}{2}B^2 g^{\mu\nu}-B^\mu B^\nu,\\ \label{10}
F^{*\mu\nu}&=&u^{\mu}B^{\nu}-u^{\nu}B^{\mu}, \\\label{11}
B^{\mu}B_{\mu}&=& B^2, \ \  B^{\mu}=\frac{1}{2}\epsilon^{\mu \nu \alpha \beta}u_{\nu}F_{\alpha \beta},\ E^{\mu}=0.
\end{eqnarray}
In the given context, the symbols $\epsilon$, P, and $F^{*\mu \nu}$ represent the energy density, pressure, and the dual tensor of the electromagnetic field, respectively. The vectors $B^{\mu}$, $E^{\mu}$ and $M^{\mu}$ correspond to the magnetic field, the electric field, and the magnetic polarization in the local rest-frame of the fluid. The metric tensor in a flat spacetime is denoted as $g_{\mu \nu}=diag \lbrace-,+,+,+\rbrace$. The four velocity of the fluid, represented by $u^{\mu}=\gamma(1, \vec v)$, satisfies the conditions $u^\mu u_\mu=-1$.

To simplify our calculation, we apply the longitudinal boost invariance along beam line, and work  with Milne coordinates rather than the standard Cartesian.

\begin{eqnarray}\label{Q3}
(\tau, x, y, \eta)=\left(\sqrt{t^2-z^2},x,y,\frac{1}{2}ln\frac{t+z}{t-z}\right).
\end{eqnarray}
Here, the metric is given by:
\begin{eqnarray}\label{Q4}
g^{\mu\nu}=diag(1,- 1,- 1,- 1/\tau^2), \ \ \ \ g_{\mu\nu}=diag(1,- 1,- 1,- \tau^2)
\end{eqnarray}
Furthermore, it is postulated that the magnetic field is positioned at a right angle to the reaction plane, aligning itself along the $y$-axis in a fluid with infinite electrical conductivity and a non-zero magnetization. The expansion of the flow in the transverse plane is solely assumed to occur in the $x$-direction. The preservation of boost invariance in the direction of the beam line ($z$-axis) enables us to confine the analysis to the $z = 0$ plane, where symmetry dictates that $u_z = 0$. Consequently, the initial configuration of our investigation can be described as follows:
\begin{eqnarray}
\label{12}
 u^\mu&=&\bar{\gamma}(1,  v_x, 0, 0) ,  \ \ B^\mu=(0, 0, B, 0),\\ \label{13}
M^\mu &=&\chi_m B^\mu   , \qquad  M \equiv \chi_m B =\sqrt{M^\mu M_\mu}. 
\end{eqnarray}
In the given scenario,  $\bar{\gamma}=\frac{1}{\sqrt{1-u_x^2}}$, and $ u_\mu B^\mu=0$ is fulfilled. Besides, in the subsequent discussion, we will make the assumption that the transverse velocity $u_x$ is non-relativistic. As a result, we will only consider terms up to the first order in $u_x$.  By applying the definition of $u^2=-1$, it can be deduced that $u^\mu = (1, u_x,0, 0)$ and $\bar{\gamma} \to 1$.

 Upon analyzing the initial setup, it can be proven that $\Omega^{\mu \nu}=0$. Moreover, Eq~(\ref{6}) exclusively consists of non-zero terms in the context of flat space-times when referring to an uncharged conformal fluid that is slightly out of equilibrium. The coefficients $\eta$, $\tau_\pi$, $\lambda_1$, $\lambda_2$, and $\lambda_3$ are commonly referred to as the first and second-order viscous transport coefficients.

Conformal second-order hydrodynamics is defined by the equations of motion
\begin{eqnarray}\label{14}
\nabla_\mu(T_{(0)F}^{\mu\nu}+T_{EM}^{\mu\nu}+\pi^{\mu\nu} )&=&0.
\end{eqnarray}
 By substituting Eqs~(\ref{5}-\ref{6}) into Eqs~(\ref{14}) and considering all the aforementioned assumptions, we arrive at the following set of coupled equations:
  \begin{eqnarray}
 \label{15}
 &&\frac{\epsilon+P+(1-\chi ) B^2}{\tau}+ \partial_x \left[u_x(\epsilon+P+(1-\chi) B^2)\right]+\frac{1}{2} \partial_\tau \left(B^2+2 \epsilon\right) \\  \nonumber 
&& +\eta f_1+\eta \tau_\pi f_2+ \lambda_1 f_3=0,
 \end{eqnarray}
  
 \begin{equation}
 \label{16}
 \begin{aligned}
&\frac{u_x(\epsilon+P+(1-\chi ) B^2)}{\tau}+ \partial_\tau \left[u_x(\epsilon+P+(1-\chi) B^2)\right] \\
&+\frac{1}{2} \partial_x \left(2 P +(1-2\chi)B^2\right) +\eta g_1+\eta \tau_\pi g_2+ \lambda_1 g_3=0,
 \end{aligned}
 \end{equation}
 here
 \begin{equation}
 \label{17}
 f_1=-\frac{4}{3 \tau^2}+\frac{4}{3 \tau} \partial_x u_x,
 \end{equation}
  \begin{equation}
 \label{18}
 f_2=-\frac{8}{9 \tau^3}+\frac{2}{3 \tau^2} \partial_x u_x - \frac{2}{9 \tau} (\partial_x u_x)^2-\frac{4}{3 \tau} \partial_\tau \partial_x u_x,
 \end{equation}
  \begin{equation}
 \label{19}
 f_3=\frac{8}{9 \tau^3}-\frac{4}{3 \tau^2} \partial_x u_x - \frac{4}{9 \tau} (\partial_x u_x)^2,
 \end{equation}
  \begin{equation}
 \label{20}
 g_1=-\frac{4}{3} \partial_x^2 u_x+\frac{2}{3 \tau} \partial_\tau u_x,
 \end{equation} 
 \begin{equation}
 \label{21}
g_2=-\frac{4}{9 \tau^3} u_x + \frac{2}{9 \tau} \partial_x^2 u_x+ \frac{8}{9 }\partial_x u_x \partial_x^2 u_x+\frac{4}{9 \tau^2} \partial_\tau u_x+\frac{4}{3} \partial_\tau \partial_x^2 u_x-\frac{2 \partial_\tau^2 u_x}{3 \tau},
 \end{equation}
 \begin{equation}
\label{22}
g_3=\frac{4}{9 \tau^3} u_x - \frac{8}{9 \tau} \partial_x^2 u_x+ \frac{16}{9 }\partial_x u_x \partial_x^2 u_x-\frac{4}{9 \tau^2} \partial_\tau u_x.
\end{equation}

The homogeneous Maxwell equation ($\nabla\mu F^{\star\mu \nu}=0$) leads to the following equation:

\begin{equation}
\label{23}
\partial_x (u_x B)+\frac {B}{\tau}+ \partial_\tau B=0.
\end{equation}

To solve the set of relativistic magnetohydrodynamics equations, it is essential to have a suitable initial condition and an appropriate Equation of State (EoS). In the context of a magnetized conformal fluid in a 4-dimensional space-time, the equilibrium equation of state is given by:
\begin{equation}\label{24}
\epsilon=\frac{1}{c_s^2} P-2 M B=3 P-2 M B.
\end{equation}
This equation is derived by setting the trace of the energy-momentum tensor equal to zero ($T^{\mu}_{\mu}=0$). It is evident that in the absence of magnetization, the conformal EoS transforms into the ultrarelativistic-fluid EoS.

In the subsequent section, we numerically solve the coupled Eqs (\ref{15}), (\ref{16}), and (\ref{23}).

\section{Numerical Calculation}
In this section, we present our numerical methodology for solving the interconnected Eqs (\ref{15}), (\ref{16}), and (\ref{23}). Given that we are dealing with a (1+1D) flow, all thermodynamic variables are reliant on the coordinates $\tau$ and $x$. Consequently, these interconnected differential equations can be effectively addressed using the method of lines (MOL), which is a numerical approach employed for solving partial differential equations (PDEs). The MOL entails discretizing one variable in one dimension and subsequently integrating the resulting semi-discrete problem as a system of ordinary differential equations (ODEs). By discretizing the partial derivatives concerning the spatial variables, we derive a system of ODEs in the temporal variable. Later on, we utilize the Mathematica software to resolve this set of ordinary differential equations (ODEs) by incorporating appropriate initial values, which will be disclosed in the subsequent section.

\subsection{The initial  conditions }
To address the interconnected partial differential equations, it is imperative to establish a collection of initial conditions.
Within our research, we examine the initial energy density and the plasma's fluid velocity profile during an early time period that must be equivalent to or surpass the point at which a hydrodynamic depiction becomes applicable. The fluid under investigation is presumed to be conformal within a four-dimensional flat space-time framework.
For the purpose of this study, we shall employ the analytical conformal solutions proposed by Gubser \cite{new3}.

The Gubser fluid velocity ($u^\mu$) is characterized by two components that are not equal to zero, specifically $u^\tau$ and $u^\perp$. These components represent the boost-invariant longitudinal  and the transverse expansion, respectively. The expressions for these components are as follows:
\begin{equation}
\label{26a}
u^{\perp}(x_\perp,\tau)=\frac{q x_\perp}{\sqrt{1+g^2(x_\perp,\tau)}}.
\end{equation}

\begin{equation}
\label{26b}
u^{\tau}(x_\perp,\tau)=\frac{1+q^2 x^2_\perp+q^2 \tau^2}{2q\tau\sqrt{1+g^2(x_\perp,\tau)}}.
\end{equation}

Subsequently, the conformal hydrodynamic Gubser's  solution  shall be employed as the initial condition at $\tau_0$. Specifically, it is postulated that the fluid adheres to the characteristics of a Gubser fluid during the initial proper time $\tau_0$. Consequently,
It is hypothesized that the initial condition of our proposed transverse fluid velocity profile at the early time $\tau_0$ is equal to Gubser's velocity solution, as denoted by:
\begin{eqnarray}
\label{25}
&& u_x(x,\tau_0) = \frac{q x}{\sqrt{1+g^2(x,\tau_0)}}, 
\nonumber \\ &&
u_{\tau}(x,\tau_0)=-1 \simeq -u^{\tau}_g(x,\tau_0)
\end{eqnarray}
The function $g(x,\tau)$ is hereby defined as follows:
\begin{equation}
\label{26}
g(x,\tau)=\frac{1+q^2 x^2-q^2 \tau^2}{2 q \tau}.
\end{equation}

Furthermore, it is posited that the fluid being studied exhibits characteristics akin to those of Gubser's hydrodynamic fluid at the specific time of $\tau=\tau_0$. As a result, the value of $\epsilon_{x}^{}(x,\tau_0)$ can be expressed as:
\begin{equation}
\label{27}
\epsilon_{x}^{}(x,\tau_0)=\epsilon_g(x,\tau_0),
\end{equation}
where, for  the inviscid case,  $\epsilon_g(x,\tau)$ is given by~\cite{new3}:
\begin{equation}
\label{28}
\epsilon_g(x,\tau)=\frac{\hat{\epsilon}_0}{\tau^{4/3 }} \frac{(2 q)^{8 / 3}}{\left[1+2 q^2\left(\tau^2+x_{}^2\right)+q^4\left(\tau^2-x_{}^2\right)^2\right]^{4 / 3}},
\end{equation}
where $\hat{\epsilon}_0$  and $q$ are  constants. The reciprocal of $q$ is directly proportional to the transverse size of the plasma.

For viscous case  $\epsilon_g(x,\tau)$ is given by:

\begin{eqnarray}
\label{29}
{\epsilon}_g(x,\tau)&=&f^* T_g^4 \\
\label{30}
T_g&=&\frac{1}{\tau f_*^{1 / 4}}\left(\frac{\hat{T}_0}{\left(1+g^2\right)^{1 / 3}}+\frac{\mathrm{H}_0 g}{\sqrt{1+g^2}}\left[1-\left(1+g^2\right)^{1 / 6}{ }_2 F_1\left(\frac{1}{2}, \frac{1}{6} ; \frac{3}{2} ;-g^2\right)\right]\right).
\end{eqnarray}

where $\hat{T}_0, \ f^* $, and $\mathrm{H}_0$ are   constant coefficients,  and $_2F_1$ denotes a hypergeometric function:

\begin{eqnarray}
\label{31}
_2F_1(\alpha,\beta;\gamma;z) &=&1+\frac{\alpha \beta}{\gamma}z+\frac{\alpha(\alpha+1)\beta(\beta+1)}{\gamma(\gamma+1)}\frac{z^2}{2}\nonumber \\ &&
+\frac{\alpha(\alpha+1)(\alpha+2)\beta(\beta+1)(\beta+2)}{\gamma(\gamma+1)(\gamma+2)}\frac{z^3}{3!}+\cdots
\end{eqnarray}

Furthermore, the magnetic field profile at the initial time ($\tau_0$) is taken into consideration as expressed by the equation:

\begin{eqnarray}
\label{32}
B^2(\tau_0,x)=\alpha f^* T_g^4(\tau_0,x).
\end{eqnarray}
In this particular scenario, the value of $\alpha$ remains constant and can be selected from the range $\alpha \in [0.1,0.7]$. The deliberate selection of the magnetic field profile at $\tau_0$ aims to simplify the numerical computation, although alternative magnetic profiles could also be considered. It is worth noting that, with this magnetic profile, the ratio of $B^2(\tau_0)$ to $\epsilon(\tau_0)$ is found to be less than 1.

By utilizing the numerical method, we are able to solve the RMHD equations in 1+1D and obtain the space-time evolution of the magnetic field, energy density, and transverse fluid velocity of the plasma ($B(\tau, x), \epsilon(\tau, x), v(\tau, x)$).  The results for these quantities are presented in the  subsection.

\subsection{Numerical results  with  magnetized conformal fluid} 
In this particular section, we will be showcasing our proposed resolutions utilizing the conformal equation of state (EoS) for fluid. To be more precise, we have successfully tackled the interconnected problems of relativistic hydrodynamics and Maxwell's equations by incorporating the conformal EoS, which is represented by the equation $\epsilon=3 P-2 M B$.

To solve the coupled relativistic hydrodynamic and Maxwell's equations for a viscous fluid, it is necessary to determine the transport coefficients involved. In the case of a conformal viscous fluid, there are five transport coefficients: $\eta$, $\tau_\pi$, $\lambda_1$, $\lambda_2$, and $\lambda_3$, which are commonly known as the first and second-order viscous transport coefficients.  These  are determined through microscopic theories. In this study, we have utilized transport coefficients calculated using two microscopic theories: Gauge/Gravity and perturbative QCD (pQCD). Table.\ref{a33} presents the results obtained from these theories. The column labeled "Gauge/Gravity" corresponds to the $N = 4$ SYM theory in the limit of infinite 't Hooft coupling. Finite coupling corrections to the gauge/gravity results have been calculated in references \cite{new4}-\cite{new9}. The column labeled "pQCD" refers to results obtained from perturbative QCD in the limit of small gauge coupling. Finite coupling corrections to some pQCD results have been calculated in references \cite{new10}-\cite{new14}. Table.\ref{a33} summarizes the current knowledge about the values for  $\eta$, $\tau_\pi$, $\lambda_1$, $\lambda_2$, and $\lambda_3$ in two different approaches.
  
 \begin{table}
 	 	\centering
 	\begin{tabular}{c|cccc}
 		\hline & Gauge/Gravity & pQCD \\
 		\hline$\epsilon(P)$ & $3 \mathrm{P}$ & $3 P $ \\
 		$\eta$ & $\frac{\epsilon+P}{4 \pi T}$  & $\frac{3.85(\epsilon+P)}{g^4 \ln \left(2.765 g^{-1}\right) T}$ \\
 		$\tau_\pi$ & $\frac{2-\ln 2}{2 \pi T}$ & $\frac{5.9 \eta}{\epsilon+P}$ & \\
 		$\lambda_1$ & $\frac{\eta}{2 \pi T}$ & $\frac{5.2 \eta^2}{\epsilon+P}$ & \\
 		$\lambda_2$ & $2\eta \tau_\pi -4\lambda_1 $ & $-2\eta \tau_\pi$ & \\
 		$\lambda_3$ & 0 &  $\frac{30(\epsilon+P)}{8 \pi^2 T^2}$ &  \\
 		& &  \\
 		\hline
 	\end{tabular}
 	\caption{Compilation of leading-order results for transport coefficients in
 		various calculation approaches}
 	\label{a33}
 \end{table}  
    
      The evaluation of viscosity, temperature evolution, and assessment of the magnetic field are dependent on a set of three interconnected Eqs (\ref{15}), (\ref{16}), and (\ref{23}). To solve the given equations, we employ  presumed values  for parameters 
      $q = 1/6.3,\hat{T}_0= 5.4, f^* = 15$ and $\mathrm{H}_0 = 0.35$  
               at proper time  $\tau=0.6$ $fm$~ \cite{new3}.
     
       Moreover, we select  a constant  value for  $B^2(\tau_0,x)/{\epsilon(\tau_0,x)}$  at $\tau_0=0.6$ fm. The initial transporter coefficients are determined by referencing Table.\ref{a33}.
        Additionally, we make use of the relationship $\epsilon=3P-2MB$, where $M=\chi_m B$, and $\chi_m$ is equal to 0.05\cite{a27}.

\begin{figure}[H]
	\centering
	
	{\includegraphics[width=.48\textwidth]{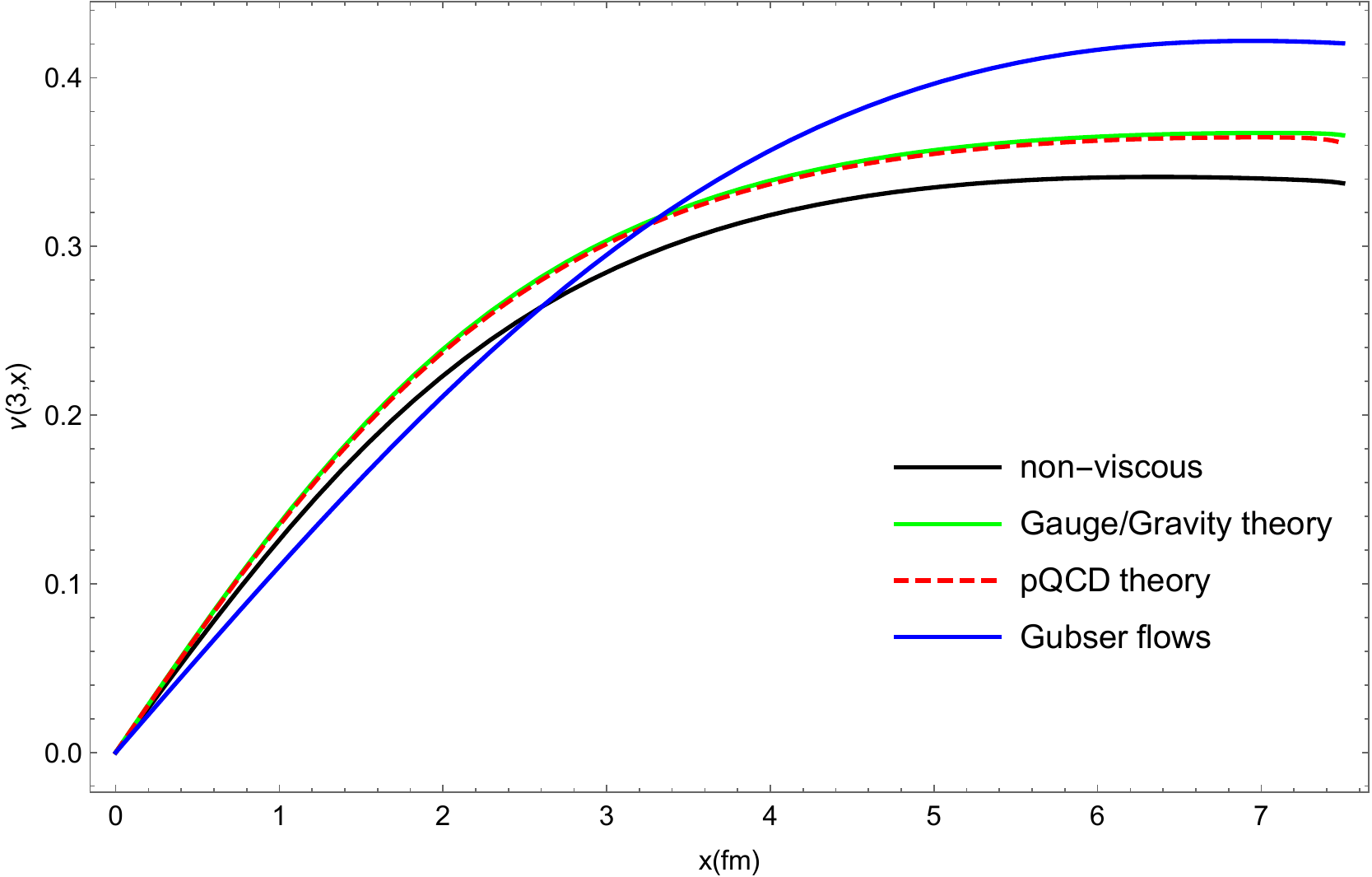}}
	{\includegraphics[width=.48\textwidth]{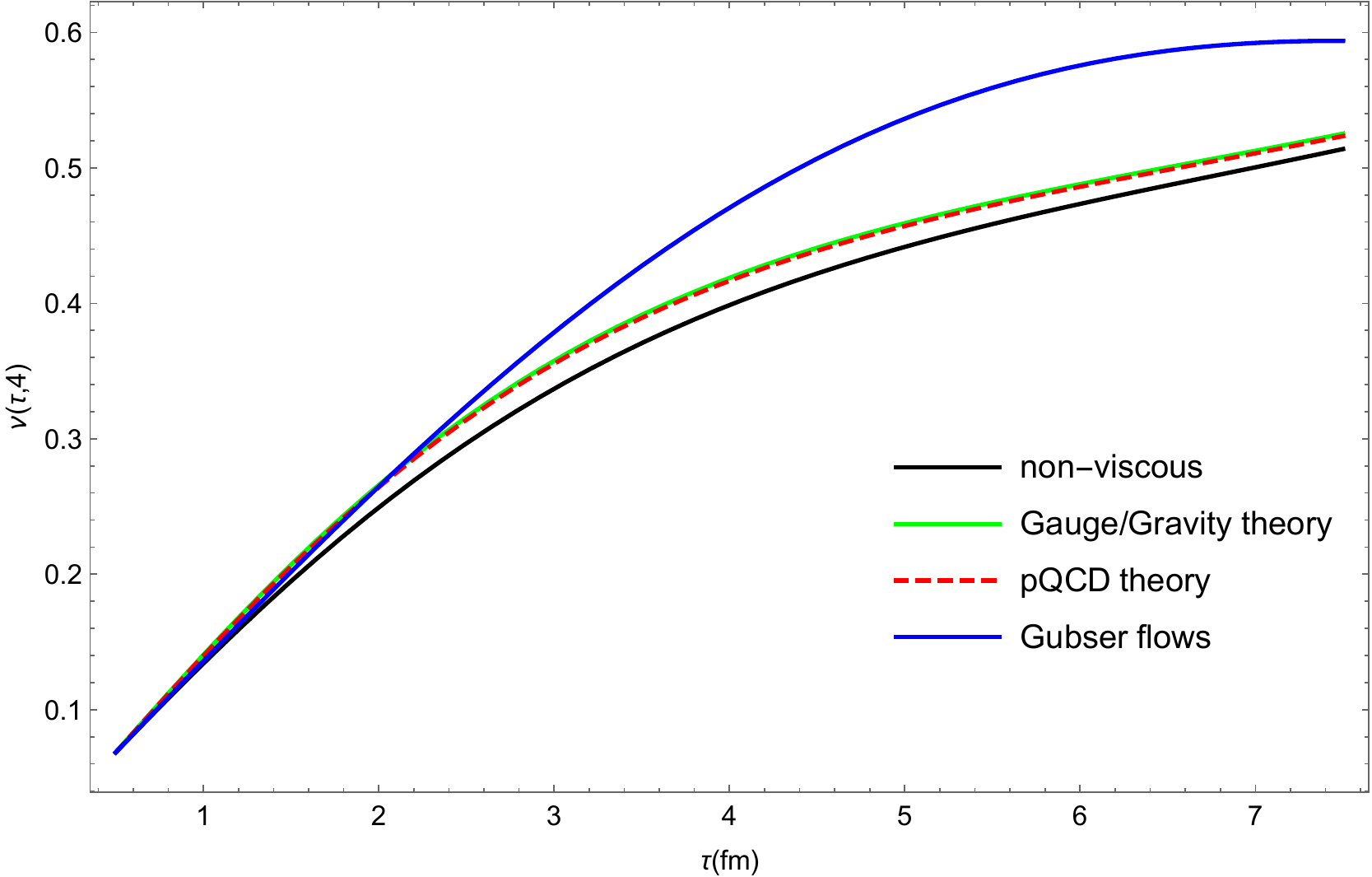}}
	\caption{Transverse velocity $v_x$ versus $x$ plotted at $\tau = 3 \ fm$ for different scenarios( left panel) corresponds to $B^2(\tau_0 , x)/ \epsilon(\tau_0 , x)=0.1$ . Transverse velocity $v_x$ versus $\tau $ plotted at $x=4 \ fm$ for different scenarios ( right panel).		 The blue solid , the green solid,   the red dashed, and the black solid  curves correspond 	to, Gubser's, viscous Gauge/Gravity, viscous pQCD approaches, and non-viscous solutions, respectively.}
	\label{f1}
\end{figure}

\begin{figure}[H]
	\centering
	
	{\includegraphics[width=.48\textwidth]{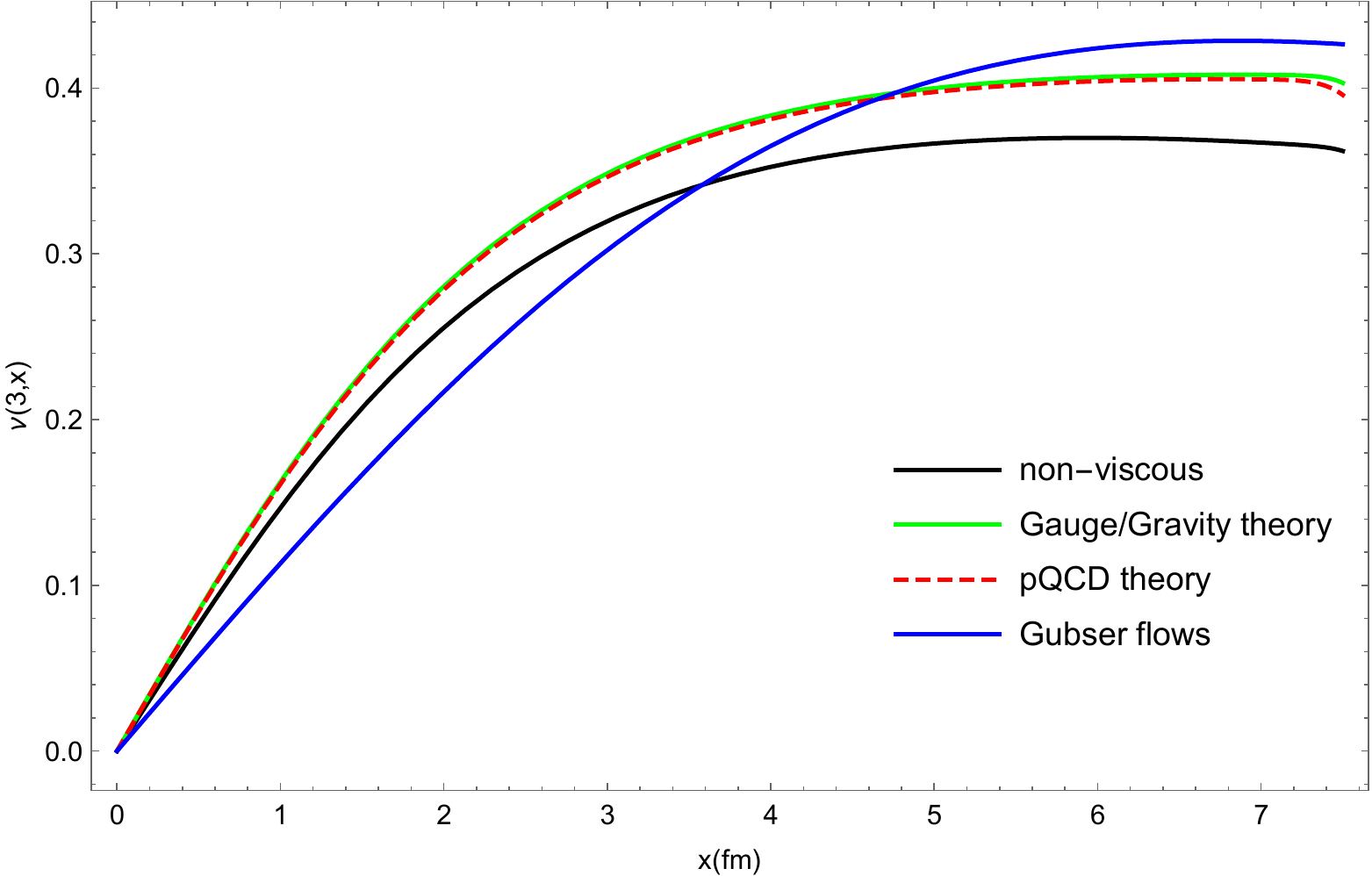}}
	{\includegraphics[width=.48\textwidth]{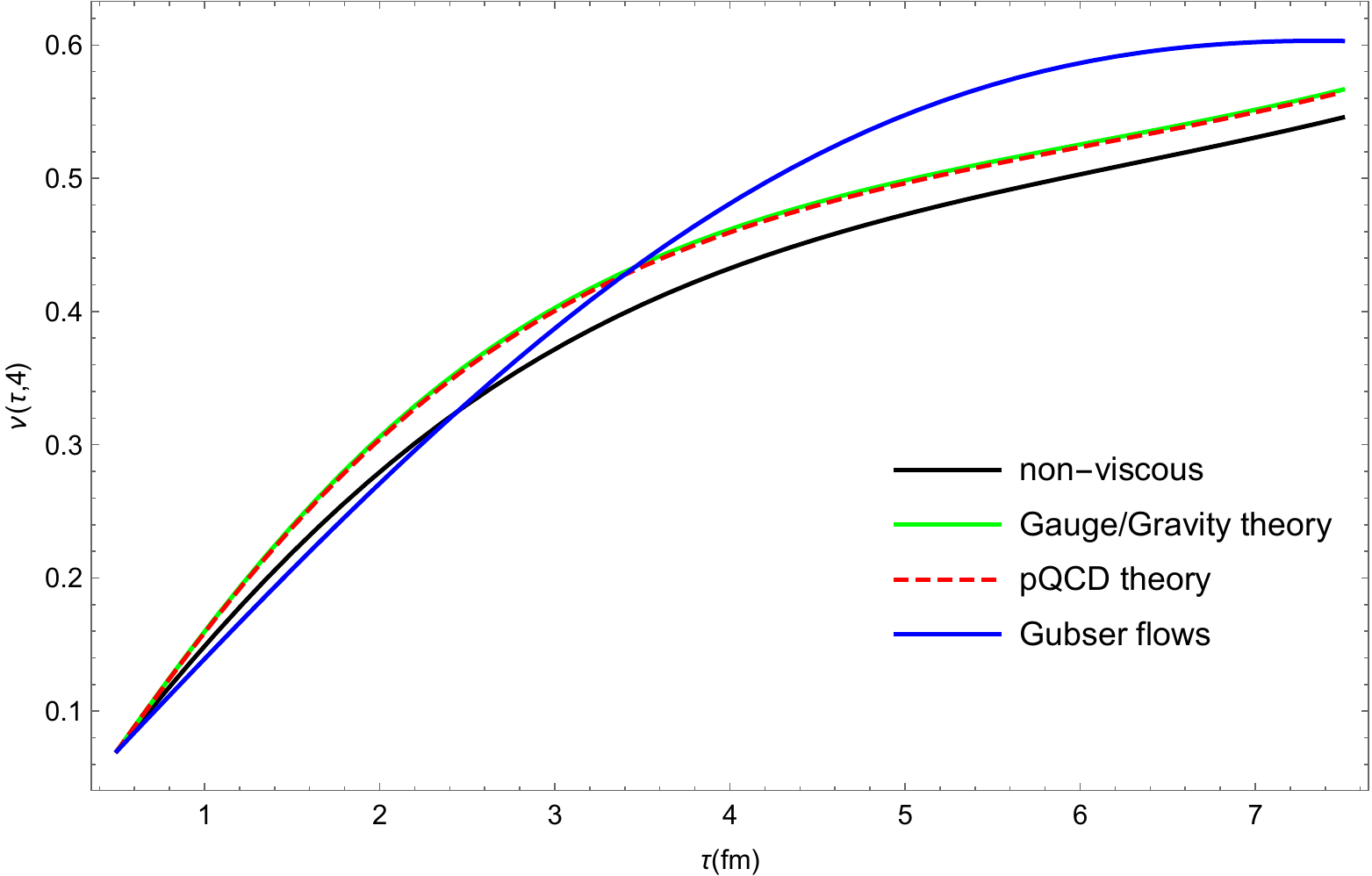}}
	\caption{Transverse velocity $v_x$ versus $x$ plotted at $\tau = 3 \ fm$ for different scenarios( left panel) corresponds to $B^2(\tau_0 , x)/ \epsilon(\tau_0 , x)=0.3$ . Transverse velocity $v_x$ versus $\tau $ plotted at $x=4 \ fm$ for different scenarios ( right panel).		 The blue solid , the green solid,   the red dashed, and the black solid  curves correspond 	to, Gubser's, viscous Gauge/Gravity, viscous pQCD approaches, and non-viscous solutions, respectively.}
	\label{f2}
\end{figure}

\begin{figure}[H]
\centering

{\includegraphics[width=.48\textwidth]{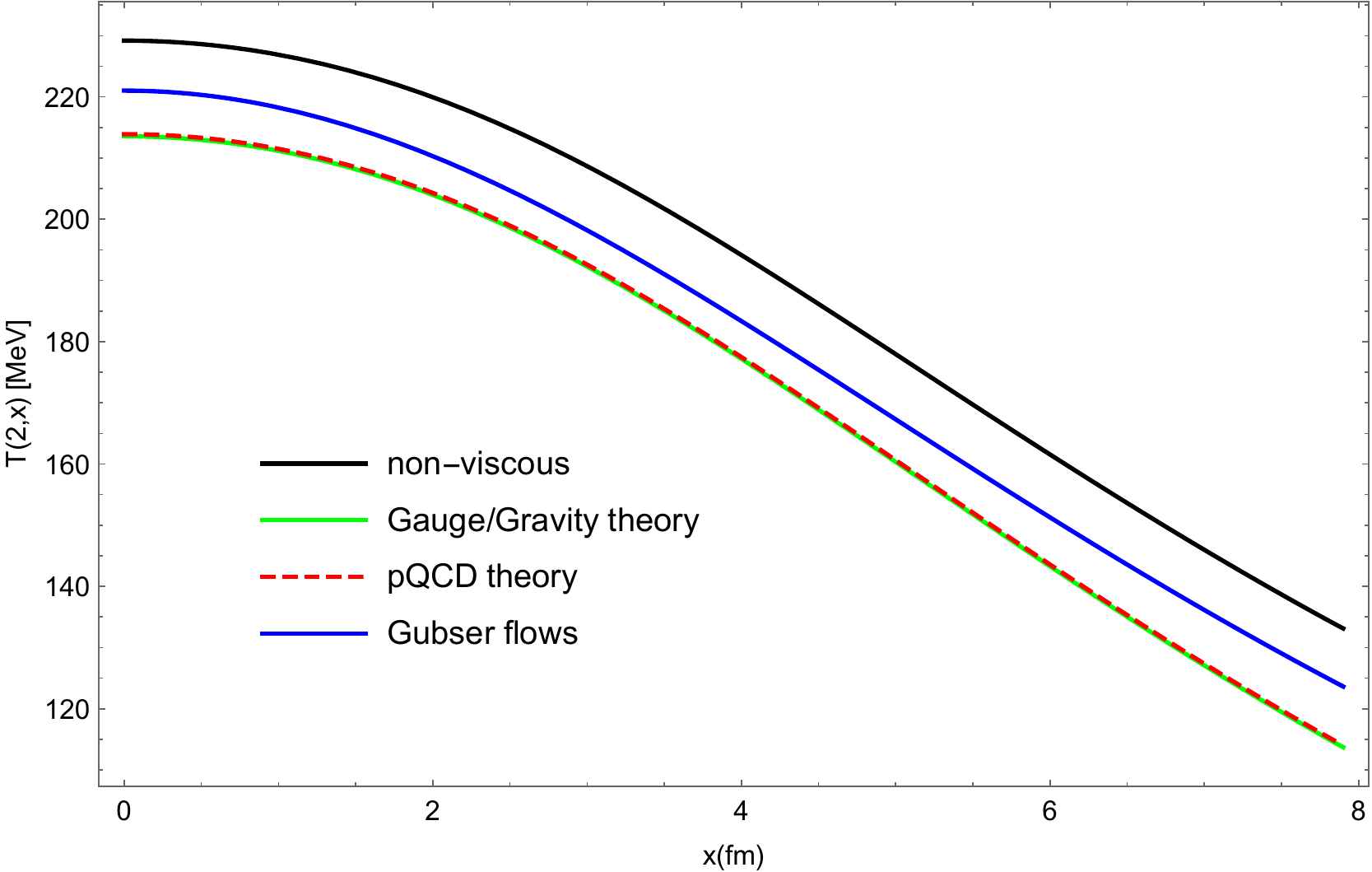}}
{\includegraphics[width=.48\textwidth]{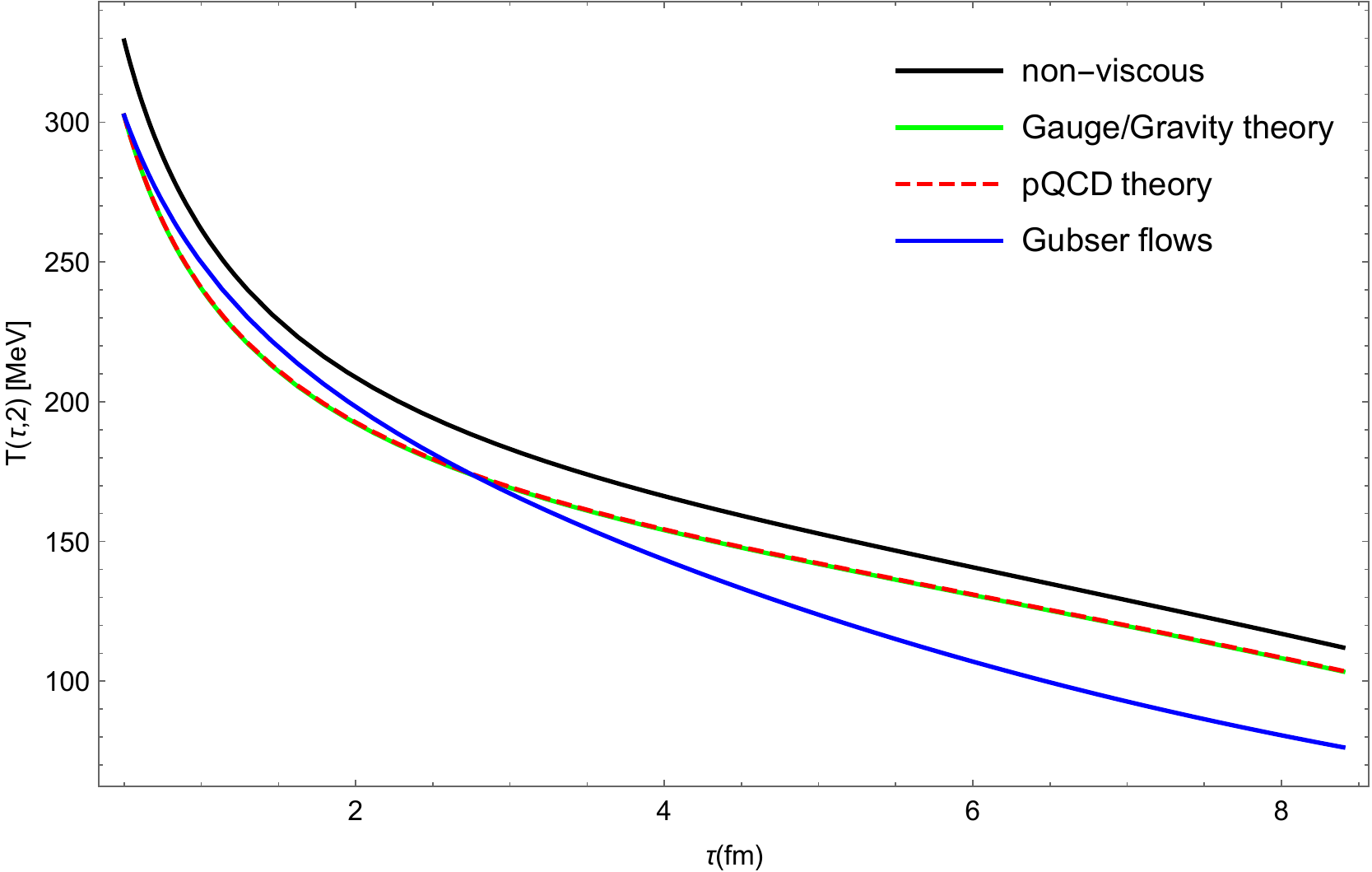}}
\caption{The local temperature $T$ versus $x$ plotted at $\tau = 2 \ fm$ for different scenarios( left panel).  $T$ versus $\tau $ plotted at $x=2 \ fm$ for different scenarios ( right panel).		 The blue solid , the green solid,   the red dashed, and the black solid  curves correspond 	to Gubser's, viscous Gauge/Gravity, viscous pQCD approaches, and non-viscous solutions, respectively.    }
\label{f3}
\end{figure}

In order to illustrate the impact of adjustments in viscosity on the development of fluid, we have graphed the transverse fluid velocity, temperature, and magnetic field with respect to either proper time $\tau$ or the spatial variable $x$. More specifically, Figs.\ref{f1}  and~\ref{f2}  show the transverse fluid velocity $v_x$ ($v_x=\frac{u^x}{u^\tau}$) as a function of $x$ or $\tau$.
The left panel of the Fig.\ref{f1} displays the transverse fluid velocity as a function of  $x$ at a fixed value of $\tau = 3 \ fm$. On the other hand, the right panel shows the transverse fluid velocity as a function of $\tau$ at a fixed value of $x = 4 \ fm$. In this diagram, we present a comparison of the solutions obtained from various models. The solid lines depicted in black, blue, and green correspond to the assessment of fluid velocity for an ideal relativistic fluid, Gubser's fluid, and a viscous relativistic fluid, respectively. The determination of the viscous coefficients in the latter case is accomplished through the Gauge/Gravity approach. Furthermore, the red dotted line illustrates the evolution of fluid velocity for a viscous relativistic fluid, where the viscous coefficients are determined using a perturbative QCD model. The intriguing aspect lies in the fluctuation of the transverse velocity of a viscous fluid, which remains consistent regardless of the specific model employed for calculating the viscous coefficients. This consistent pattern is clearly evident in  Figs. \ref{f2}  \ref{f3}. Nevertheless, within the realm of a viscous fluid, the velocity component $v_x$ demonstrates a rise in contrast to the velocity progression of the non-viscous fluid. In Fig.~\\ref{f2},   the transverse fluid velocity as a function of $x$  at fixed $\tau$ (the left panel) or as a function of $\tau$ at fixed $x$ (the right panel) has been plotted for an initial value of  $B^2(\tau_0,x)/{\epsilon(\tau_0,x)}=0.3$. 
The evident impact of increasing the initial magnetic field is reflected in the amplified discrepancy between the velocity component $v_x$ of the viscous fluid and that of the non-viscous fluid.

The Fig.\ref{f3} shows 
the local temperature $T$ as a
function of $x$ for selected value of $\tau=2$ $fm$ (left panel), or as a function of $\tau$ for selected value of $x=2$ $fm$ (right panel) for an ideal fluid (black solid curve), Gubser's fluid (blue solid curve), and non-zero viscosity (green solid and  red  dotted curves). The cooling rate is faster for a viscous  fluid compared to an ideal fluid, which contradicts the findings mentioned in Ref.~\cite{new14}.

\begin{figure}[H]
\centering

{\includegraphics[width=.48\textwidth]{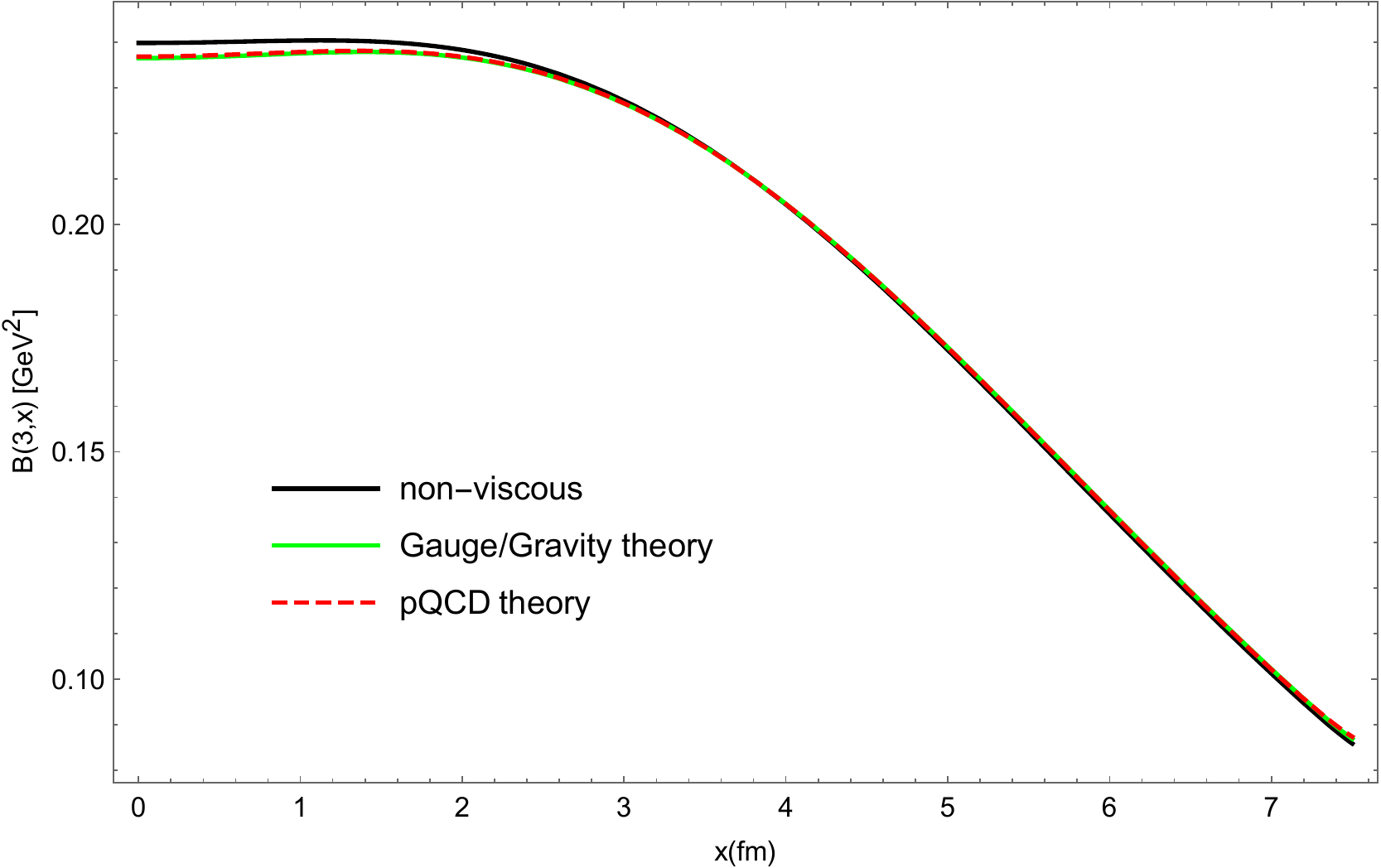}}
{\includegraphics[width=.48\textwidth]{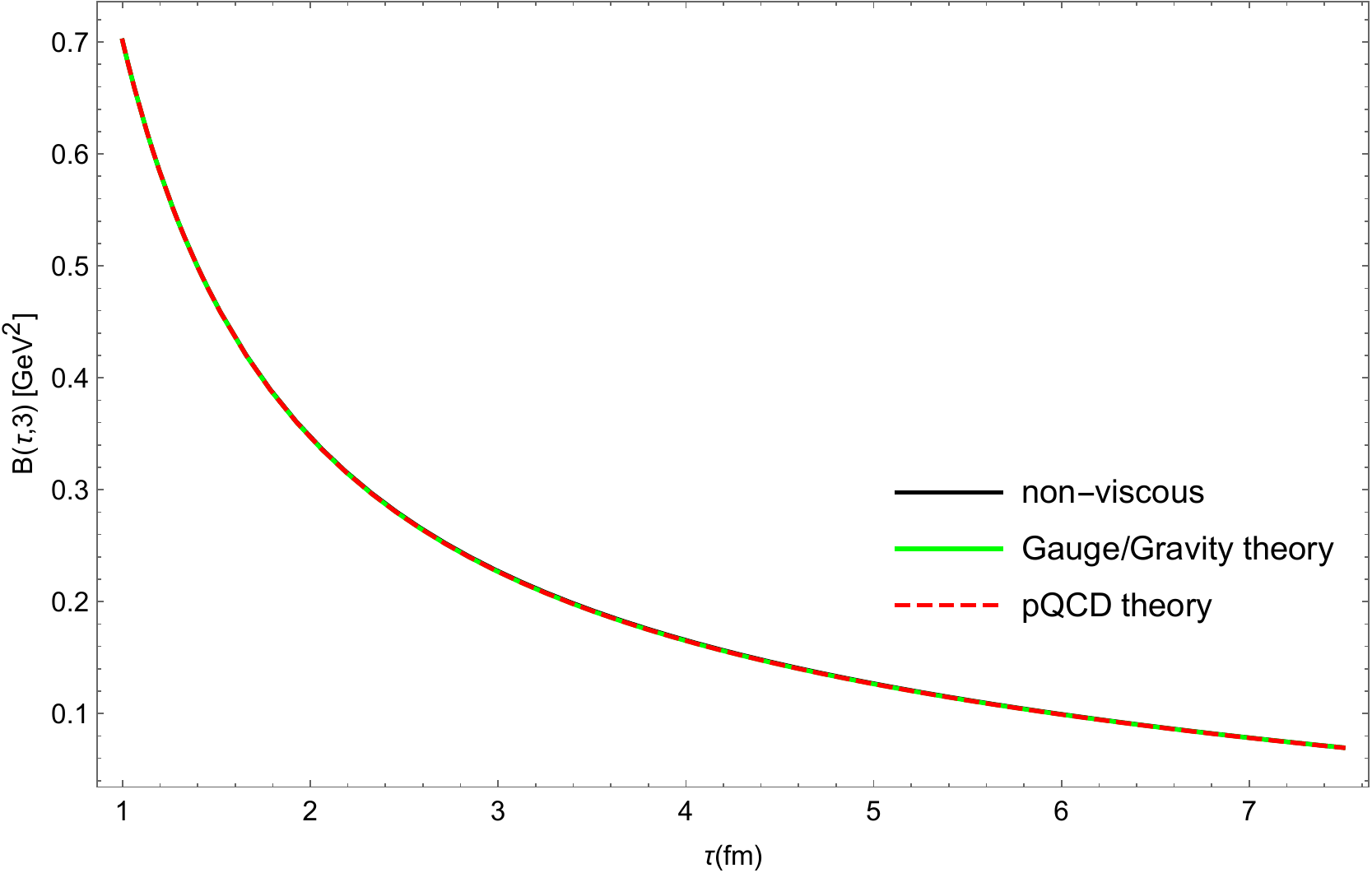}}
\caption{  The magnetic field  $B$ versus $x$ plotted at $\tau = 3 \ fm$ for different scenarios( left panel).  $T$ versus $\tau $ plotted at $x=3 \ fm$ for different scenarios ( right panel).		 The blue solid , the green solid,   the red dashed, and the black solid  curves correspond 	to Gubser's, viscous Gauge/Gravity, viscous pQCD approaches, and non-viscous solutions, respectively.}
\label{f4}
\end{figure}

 Fig.~\ref{f4} presents a graphical representation of the magnetic field's variation in relation to the spatial coordinate, $x$, at a fixed time, $\tau=3 \ fm$ (left panel), and in relation to the temporal coordinate, $\tau$, at a fixed spatial coordinate, $x=3 \ fm$ (right panel). The curves depicting the viscose fluid are shown as red dotted and green solid lines, while the ideal fluid is represented by a black solid line. It is important to note that during the initial stages, there is a minor impact on the assessment of the magnetic field. However, our research findings indicate that the viscosity of the fluid has an insignificant influence on the temporal and spatial evolution of the magnetic field. This conclusion remains valid, at least for the specific values of magnetic susceptibility and initial magnetic field chosen for this study.

\begin{figure}[H]
\centering

{\includegraphics[width=.48\textwidth]{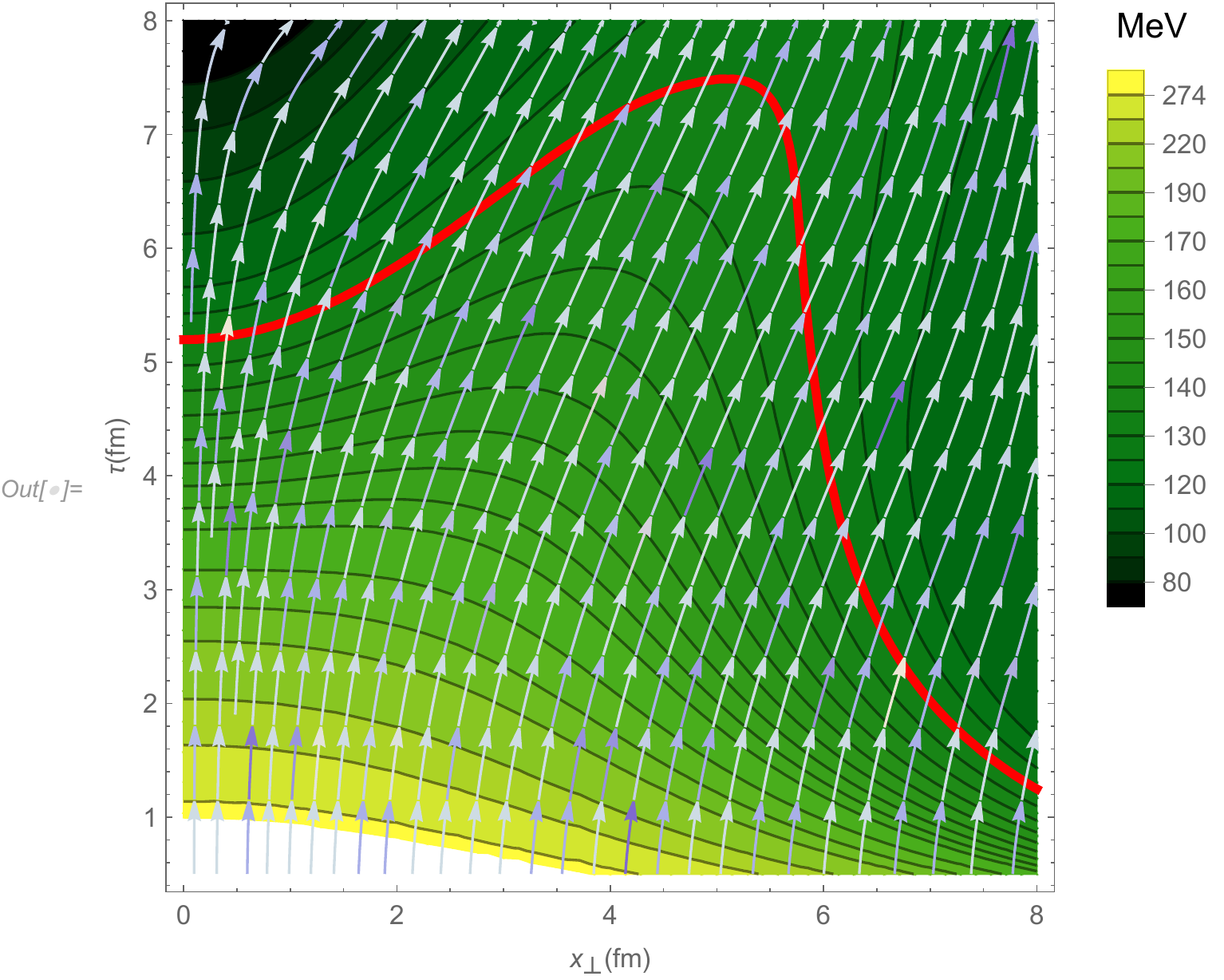}}
{\includegraphics[width=.48\textwidth]{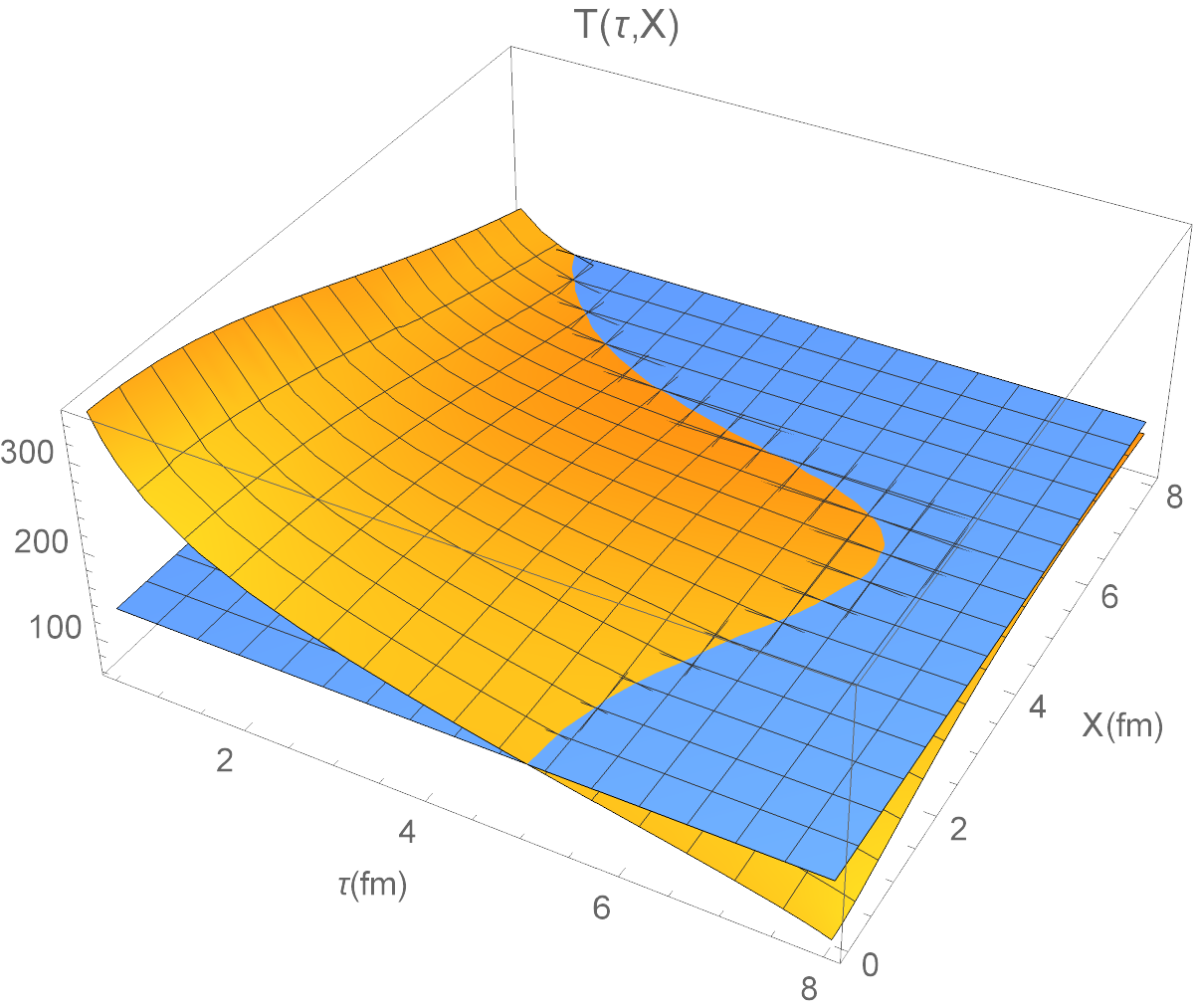}}
\caption{The two-dimensional  fluid velocity $(u^\tau/\sqrt{u_\tau^2+u_x^2},u^x/\sqrt{u_\tau^2+u_x^2})$ is plotted with parameters chosen as $q=1/6.3(fm)^{-1}$, $\tau_0=0.6 (fm)$, $\hat{T}_0=5.4$ ${GeV}/{fm^3}$, In our calculations, $u_\tau=1$, and $v_x=u^x$(Left panel). 
		The local temperature  
		$T(\tau,x)$ 
		as a  function of transverse coordinate  $x$  at $\tau=3 \ fm$   (Right panel).}
\label{f5}
\end{figure}

We also  present the two-dimensional fluid velocity and the contours of constant temperature in Fig.~\ref{f5}. The plot has been exhibited for values of $(\tau,x)$ wherein $\epsilon$ is positive. The prominently delineated black contour corresponds to the constant temperature.

 In summary, the findings suggest that the viscosity  dose  considerable  impact on the evolution of  the transverse fluid velocity and local temperature.

Investigating the impact of magnetic field and magnetization on the anisotropic longitudinal and transverse pressures of a viscous fluid is an intriguing area of study. In the local rest frame, the ratio of longitudinal pressure to transverse pressure, represented as $\frac{P_L}{P_T}$, is referred to as the effective pressure anisotropy. To analyze this phenomenon, it is advantageous to acknowledge the high level of symmetry exhibited by the Bjorken flow in a conformal system. This symmetry enables us to express the effective pressures, along with related quantities such as the anisotropy, in terms of the energy density and its derivatives. Specifically, 
implying 
 the conditions $\nabla_\mu T^{\mu \nu}=0$ and $T_\mu^\mu=0$  for conformal Bjorken flow one obtains\cite{new15}:
\begin{equation}
\tau \partial_\tau T_{\tau \tau}+T_{\tau \tau}+\frac{T_{\eta \eta}}{\tau^2}=0, \quad-T_{\tau \tau}+2 T_{x x}+\frac{T_{\eta \eta}}{\tau^2}=0,
\end{equation}
which using $T_{\tau \tau}=\epsilon(\tau,x)$ may be recast as $P_{\mathrm{eff}}^{(\xi)}=T_{\eta}^{\eta}=-\epsilon(\tau,x)-\tau \partial_\tau \epsilon(\tau,x)$ and $P_{\mathrm{eff}}^{(x)}=T_x^x=\epsilon+\frac{\tau}{2} \partial_\tau \epsilon$. Hence, the expression for the pressure anisotropy $P_{\mathrm{eff}}^{(\xi)} / P_{\mathrm{eff}}^{(x)}$ in the case of conformal Bjorken flow can be written as follows:
\begin{equation}
\label{24b}
\begin{aligned}
&\frac{P_L}{P_T}=\frac{-1-\tau \partial_\tau \ln \epsilon}{1+\frac{\tau}{2} \partial_\tau \ln \epsilon} \simeq 1-\frac{8 C_\eta}{\tau T}+\frac{16 C_\eta\left(4 C_\eta-C_\pi\left(1-C_\lambda\right)\right)}{3 \tau^2 T^2}\\
&-\frac{9}{2 \tau} \frac{B^2}{\epsilon} \chi+\frac{24 C_\eta}{\tau T}\frac{B^2}{\epsilon} \chi-\frac{16 C_\eta\left(6 C_\eta-C_\pi\left(1-C_\lambda\right)\right)}{ \tau^2 T^2}\frac{B^2}{\epsilon} \chi+\mathcal{O}\left((\tau T)^{-3}\right).
\end{aligned}
\end{equation}
  It is customary to introduce the dimensionless combinations $C_\eta;C_\pi;C_\lambda$ (which are constants for
  conformal systems) through:
 \begin{equation}
\frac{\eta}{\epsilon}=\frac{4}{3} C_\eta T^{-1}, \quad \frac{\eta \tau_\pi}{\epsilon}=\frac{4}{3} C_\eta C_\pi T^{-2}, \quad \frac{\lambda_1}{\epsilon}=\frac{4}{3} C_\eta C_\pi C_\lambda T^{-2}.
\end{equation}
 \begin{figure}[H]
	\centering
		{\includegraphics[width=.48\textwidth]{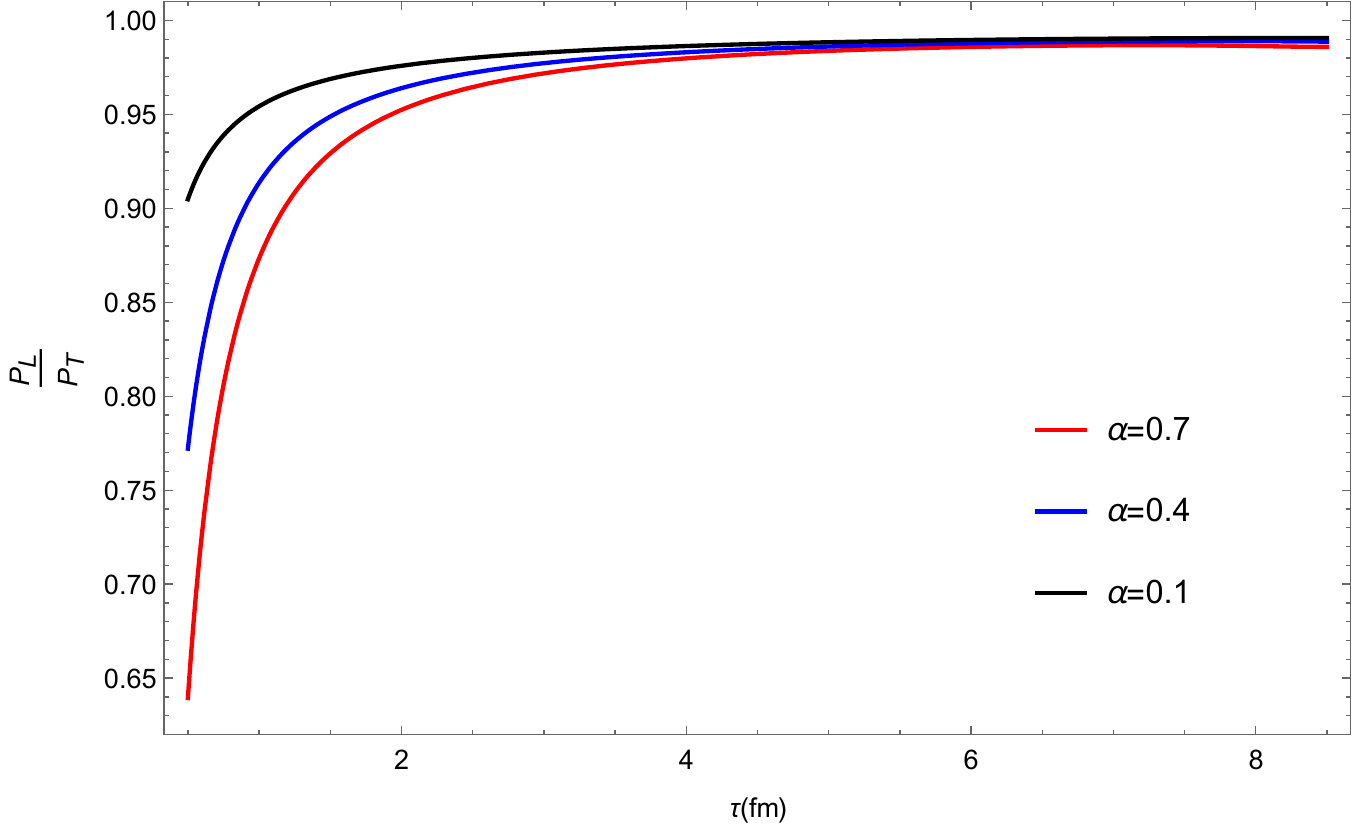}}
	\caption{Evolution of  $\frac{P_L}{P_T}$ as a functions of proper time $\tau$ for three values of $B^2(\tau_0,x)/\epsilon(\tau_0,x)$.}
	\label{f6}
\end{figure}

From (\ref{24b}), it can be observed that the inclusion of viscous and magnetic field corrections results in anisotropic longitudinal and transverse pressures.
However, as time progresses, the fluid eventually becomes isotropic. To examine the impact of magnetization and the magnetic field on the fluid's anisotropy, we present the ratio $\frac{P_L}{P_T}$ as a function of proper time $\tau$ for three distinct values of $\frac{B^2(\tau_0,x)}{\epsilon(\tau_0,x)}$. It is evident that at early times $\tau<2 \ fm$, the medium exhibits anisotropic behavior. Furthermore, selecting a larger value for $\frac{B^2(\tau_0,x)}{\epsilon(\tau_0,x)}$ leads to a greater degree of anisotropy in the fluid.

\subsection{ The spectra }

In this section, we will examine the impact of viscous terms and the application of initial condition \ref{18}, \ref{19} on the particle spectrum. It is important to note that our model is highly simplistic, as it describes a fluid in a 1+1D Bjorken scenario. Therefore, it can be considered more of a toy model. Nonetheless, we make the assumption that $x=r$, where $r$ represents the radial coordinate. Subsequently, we proceed to calculate the spectra of hadrons  and compare them with experimental data. It is evident that the model's simplicity allows for a straightforward comparison of its results with the experimental data. Despite this simplicity, the comparison yields intriguing findings.

To acquire the hadron spectra, we employ the Cooper-Frye freeze-out prescription across the freeze-out surface~\cite{cooper}.

\begin{eqnarray}
S=E\frac{d^3N}{dp^3}=\frac{dN}{p_Tdp_T dyd\varphi}=\int d\Sigma_\mu p^\mu \exp(\frac{-p^\mu u_\mu}{T_f})
\end{eqnarray} 
The temperature at the freeze out surface, denoted as $T_f$, is determined by the 4-velocity of the fluid, represented by $u^{\mu}$. The element area on the isothermal freeze out surface in space-time is given by $d\Sigma_\mu$. The freeze out surface is defined as the region where the temperature of the fluid is proportional to the energy density, expressed as $T\propto \epsilon^{1/4}$. This condition is satisfied when $T(\tau, r)=T_f$. Furthermore, within our convention, the area element that is perpendicular to the freeze out surface can be expressed as follows:
\begin{eqnarray}
d\Sigma_\mu&=&(-1, R_f, 0, 0)\tau_f r dr d\varphi  d\eta.
\end{eqnarray}
Thus, the integration measure in the Cooper-Frye formula is: 
\begin{eqnarray}
d\Sigma_\mu p^\mu&=&[-m_T\cosh(Y-\eta)+p_T R_f\cos(\varphi_p-\varphi)]\tau_f  dr d\varphi d\eta.
\end{eqnarray}
Moreover, the scalar product $p^{\mu}u_{\mu}$ within the Cooper-Frye formula is yet to be determined.
\begin{eqnarray}
p^\mu u_\mu&=&-m_T\cosh(Y-\eta)u_\tau+p_T\cos(\varphi_p-\varphi)u_r,
\end{eqnarray}
where $R_f\equiv -\frac{\partial \tau}{\partial r}=\frac{\partial_r T}{\partial_\tau T}\mid_{T_f}$.
Here $\tau=\sqrt{t^2-z^2}$, $ r$, and  $\eta=\frac{1}{2}\log\frac{t+z}{t-z}$,  are   the longitudinal proper time,  the transverse
(cylindrical) radius, and the longitudinal rapidity
(hyperbolic arc angle),  respectively. Likewise, $u_r$ represents the transverse velocity of flow while $\varphi$ denotes its azimuthal angle. On the other hand, $\varphi_p$ signifies the azimuthal angle in momentum space. Additionally, $p_T$, $m_T=\sqrt{m^2+p_T^2}$, and $Y$ refer to the measured transverse momentum, the corresponding transverse mass, and the observed longitudinal rapidity respectively. Consequently, the CF formula can be expressed as follows:
\begin{eqnarray}
S=\frac{g_i}{2\pi^2}\int_0^{x_f}\  r\ \tau_f(r)\ dr\
\Big[m_T K_1(\frac{m_Tu_\tau}{T_f})I_0(\frac{m_Tu_r}{T_f})+p_T R_f
K_0(\frac{m_Tu_\tau}{T_f})I_1(\frac{m_Tu_r}{T_f})\Big].
\label{spectrum}
\end{eqnarray}
The solution of $T(\tau_f, r)=T_f$ is denoted as $\tau_f(r)$, while the degeneracy factor for particles is represented as $g_i$.
The numerical evaluation of the integral over $r$ on the freeze-out surface yields the spectra of hadrons as a function of $p_T$. The figures present the results for the spectra of charged particles.

\begin{figure}[H]
	\centering
	
	{\includegraphics[width=.32\textwidth]{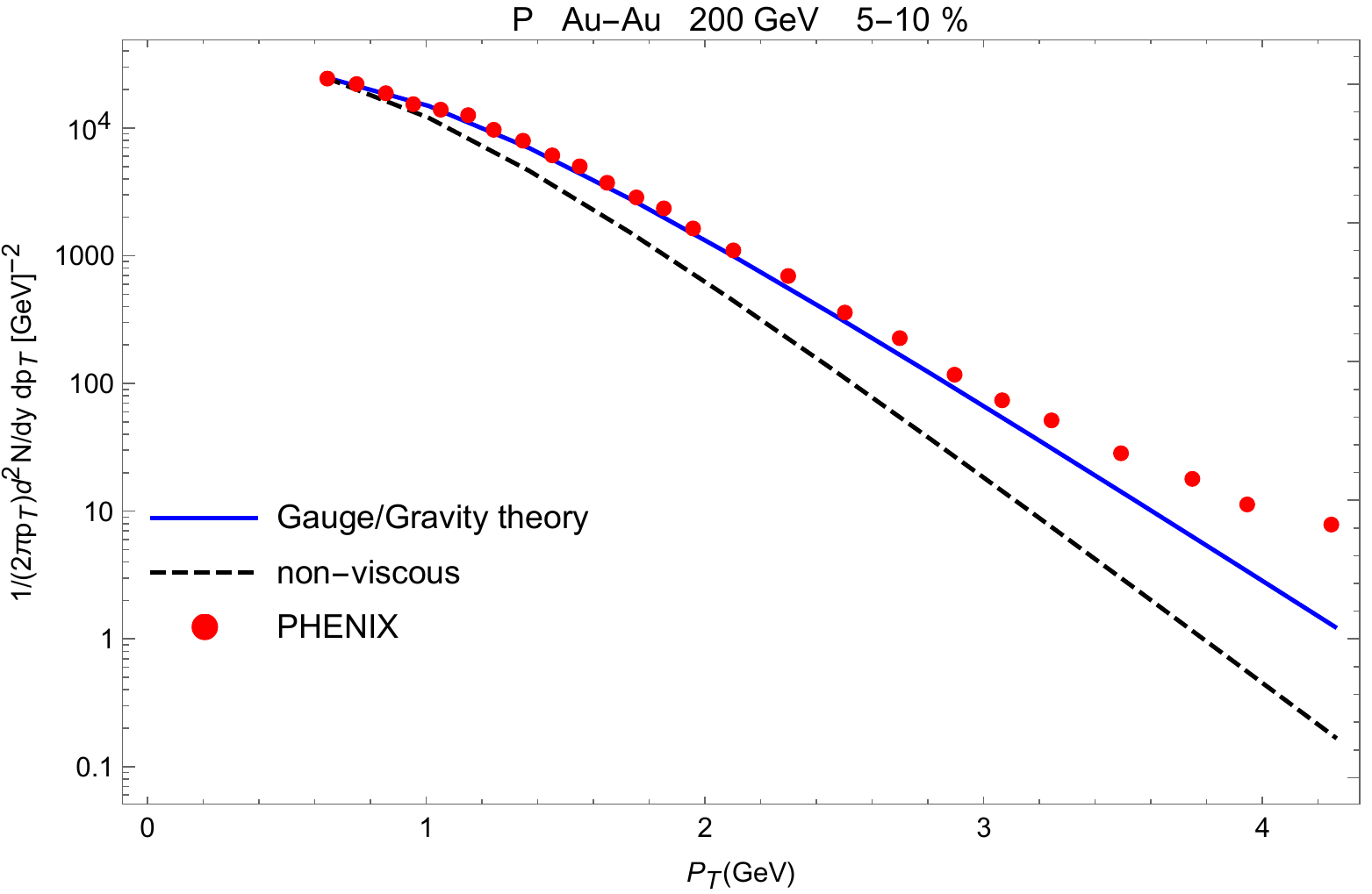}}
	{\includegraphics[width=.32\textwidth]{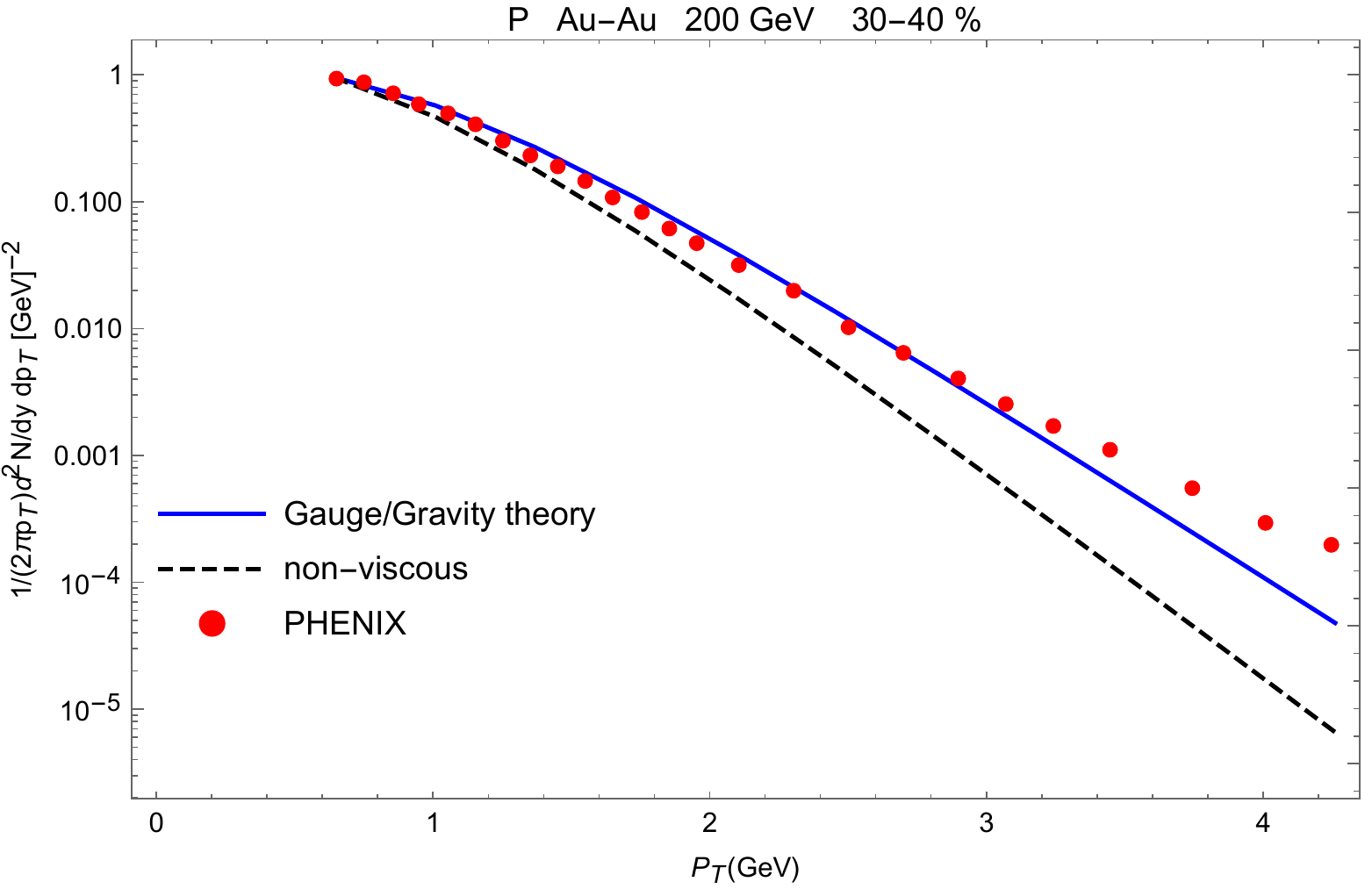}}
	{\includegraphics[width=.32\textwidth]{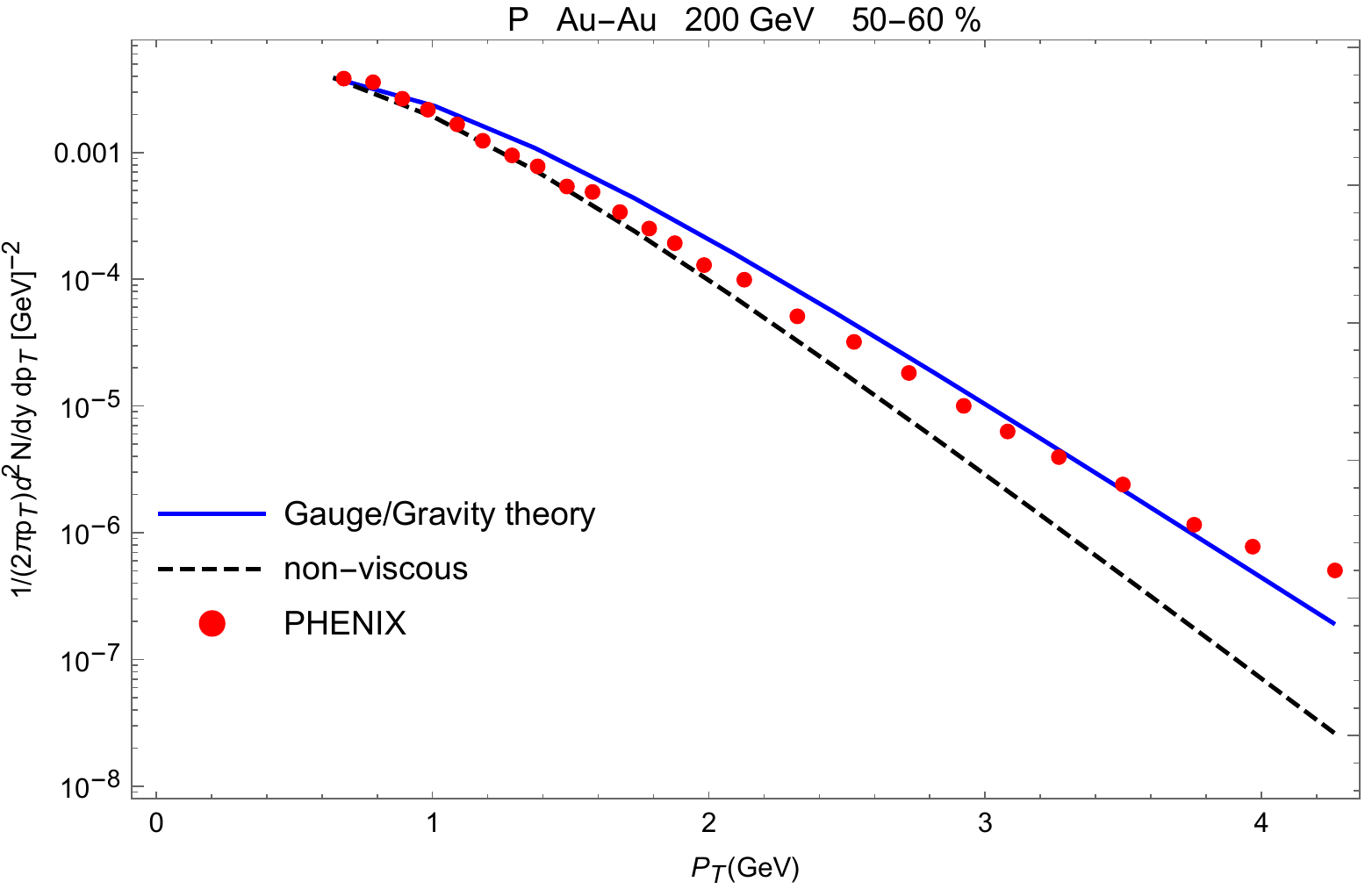}}
	\caption{Proton transverse spectrum from  Au-Au collisions for $B^2(\tau_0 , x)/ \epsilon(\tau_0 , x)=0.3$.
		The  red dot line corresponds to  PHENIX data \cite{phenix}}
	\label{f7}
\end{figure}

The hadron spectrum for the proton is illustrated in Fig.\ref{f7}. We have selected the isotherm at $T = 130 \ MeV$ as the freezeout surface for heavy ion collisions at RHIC, considering three distinct centrality classes. The proton's spectrum comparison, in terms of transverse momentum $p_T$, is examined in this study. The analysis includes   our own hydrodynamic solutions, and experimental data obtained from PHENIX ~\cite{phenix}. The curves represented by the black dashed, blue solid,  and reds dot correspond to the non-viscous solution, viscous solution, experimental results,  respectively.
It is evident that the spectra obtained from  non-viscous solutions, using the specified parameters, are not entirely satisfactory. Nevertheless, our viscous solutions exhibit a remarkable level of agreement with the experimental data obtained at PHENIX~\cite{phenix}.  
 Additionally, we observe that our viscous solutions exhibit a more accurate correspondence with the experimental findings when considering a non-central collisions, as depicted at the bottom of Fig.\ref{f7}. This is particularly evident when analyzing the chosen centrality class for heavy ion collisions within the range of $50-60\%$.
  
\begin{figure}[H]
	\centering
	
	{\includegraphics[width=.32\textwidth]{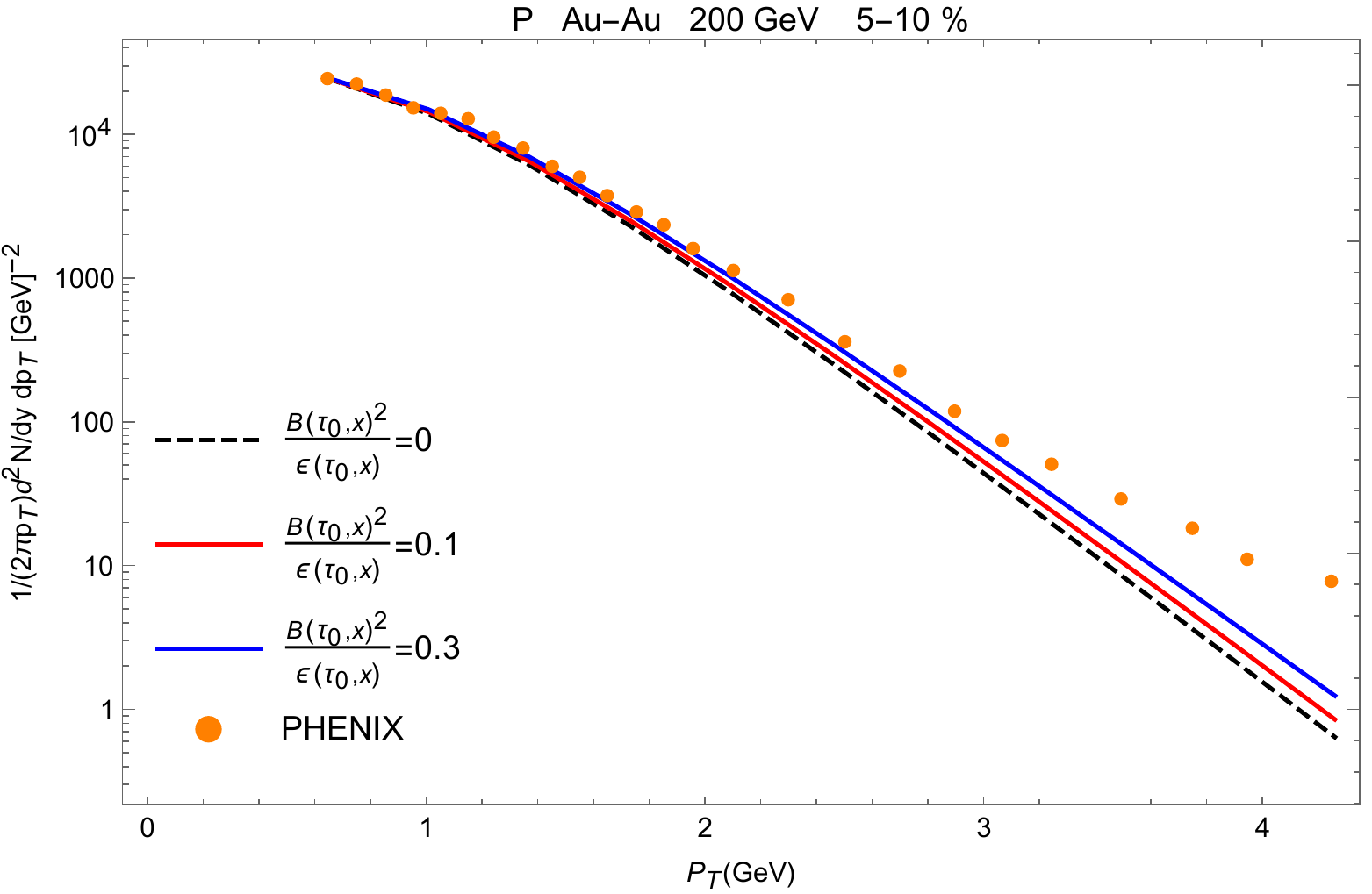}}
	{\includegraphics[width=.32\textwidth]{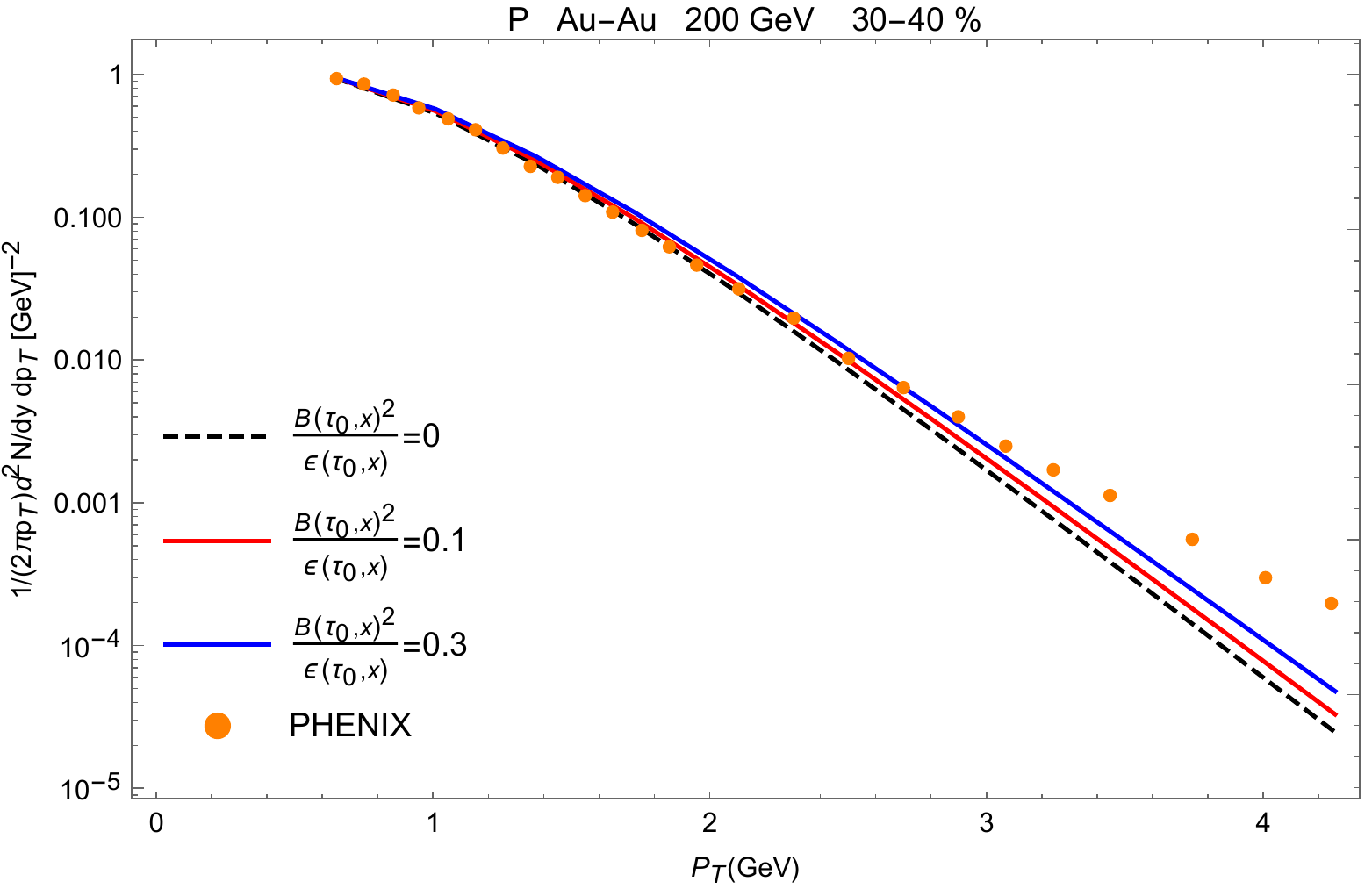}}
	{\includegraphics[width=.32\textwidth]{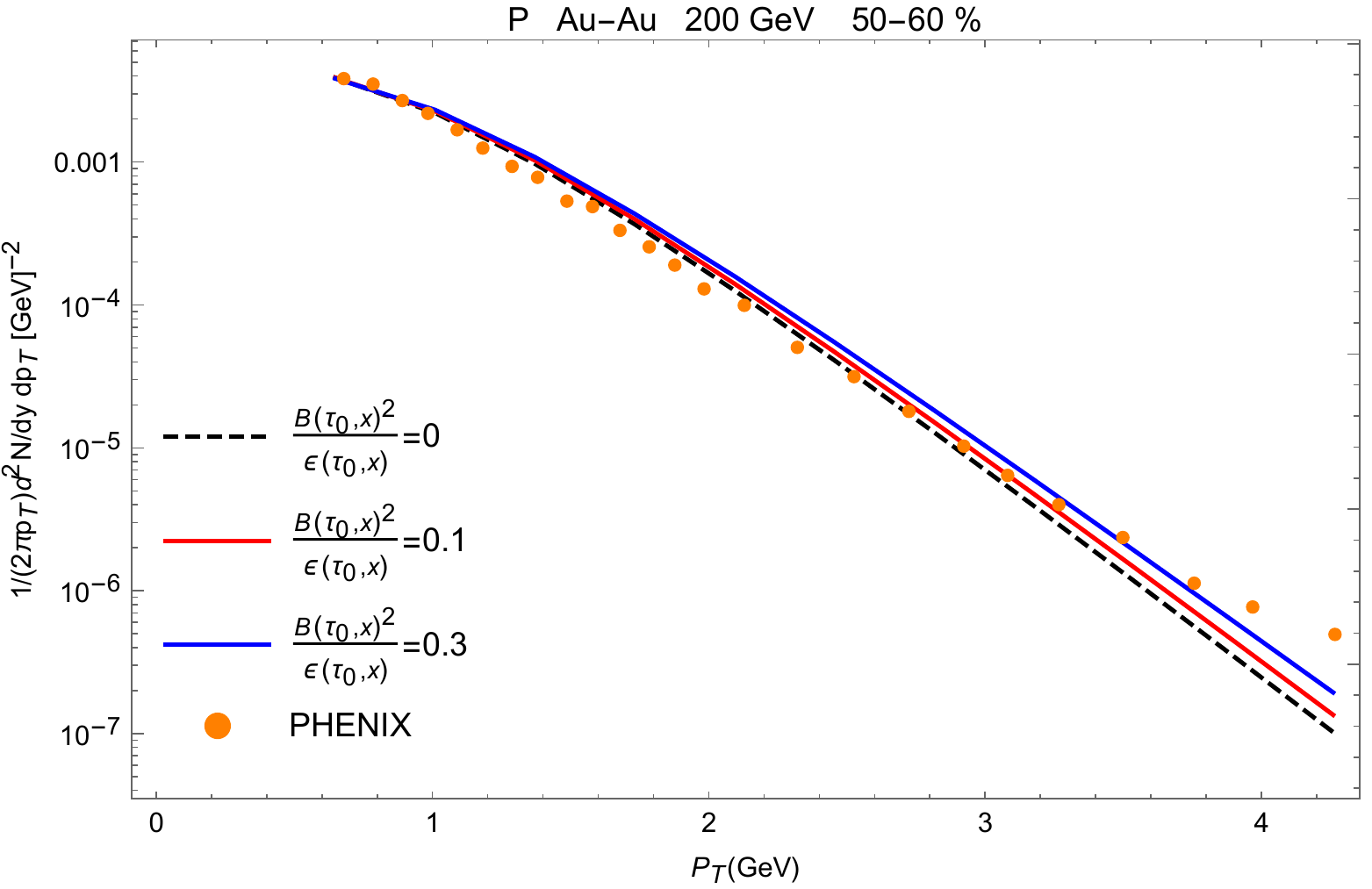}}
	\caption{Proton transverse spectrum from  Au-Au collisions for  three different values of $B^2(\tau_0 , x)/ \epsilon(\tau_0 , x)$. The dot black, the red solid, and blue solid curves are corespondent to the viscous solutions with the initial values of $\frac{B^2(\tau_0,x)}{\epsilon(\tau_0,x)} =0, 0.1, 0.3 $, respectively.		
			The  red dot line corresponds to  PHENIX data \cite{phenix}}
	\label{f8}
\end{figure}

 Furthermore, to examine the impact of the magnetic field on the proton spectrum, we have graphed the proton spectrum for three distinct initial condition values of the ratio $\frac{B^2(\tau_0,x)}{\epsilon(\tau_0,x)}$ in Fig.\ref{f8}. The black dashed, red solid, and blue solid curves correspond to viscous solutions with the ratio $\frac{B^2(\tau_0,x)}{\epsilon(\tau_0,x)}=0, 0.1, 0.3$, respectively.    
   The experimental data is represented by the orange dot curve. Our analysis reveals that the spectra of our viscous solutions once again exhibit a striking concurrence with the experimental data.      
    The provided figures, labeled as Fig.\ref{f8}, present also a comparison of results obtained for centrality classes in heavy ion collisions. The left panel represents the results within the centrality range of $5-10\%$, the middle panel represents the range of $30-40\%$, and the right panel represents the range of  $50-60\%$. 
       The findings shows that influence of the initial magnetic field on the proton spectrum is negligible.

\section{Conclusion}
In our current study, we have examined heavy ion collisions by employing a viscous Hydrodynamics model. Additionally, we have taken into account the magnetization of the matter produced during the collision. Furthermore, we have assumed the presence of a robust magnetic field, particularly during the early stages of Quark-Gluon Plasma (QGP) formation. Consequently, we have considered the influence of the magnetic field, magnetization, and viscous effects when evaluating the properties of the matter.

We successfully solved the coupled conservation and Maxwell's equations and obtained numerical results for the fluid velocity, energy density, and magnetic field. Our work is based on the (1+1D) dimensional MHD model, where the transverse magnetic field, fluid velocity, and energy density are functions of one spatial variable $(x)$ and one temporal variable $(\tau)$. Additionally, the magnetic field is oriented along the orthogonal $y$ direction. In our setup, the medium exhibits boost invariance along the $z$ direction. We made the assumption that the transverse velocity remains relatively small at all times and for various parameterizations. Therefore, we treated the transverse flow using the non-relativistic approximation and neglected the terms involving $u_x^2$ in the coupled equations. The core of our method consists of two main aspects: (1) utilizing Gubser's solutions as initial conditions for the transverse velocity and energy density, and (2) considering the QGP as a conformal viscous fluid.

In this study, the three coupled differential equations resulting from the reduction of the energy conservation, Euler equations, and Maxwell's equations were solved numerically. The solutions obtained from the numerical calculations were based on specific initial conditions. These initial conditions assumed values for parameters such as $q = 1/6.3$, $\hat{T}_0 = 5.4$, $f^* = 15$, and $\mathrm{H}_0 = 0.35$ at a proper time of $\tau = 0.6$ $fm$~ \cite{new3}. Furthermore, the ratio of the square of the magnetic field to the fluid energy density, $B^2(\tau_0,x)/\epsilon(\tau_0,x)$, was kept constant.
Two different models, namely the Gauge/Gravity and perturbation QCD models, were utilized to obtain the viscous transport coefficients $\eta$, $\tau_\pi$, $\lambda_1$, and $\lambda_2$. By employing the initial conditions and transport coefficients obtained from these models, the coupled equations were solved, and the resulting solutions were compared with both the non-viscous case and Gubser's solutions.

We have also  investigated  the effect of magnetic field   on the the effective pressure anisotropy ($\frac{P_L}{P_T} $)  of bulk matter in  high-energy heavy-ion collisions, within a (1+1)-dimensional ideal hydrodynamic model. Our findings indicate that the effect  is sensitive to the ratio of  $\frac{B^2(\tau_0,x)}{\epsilon(\tau_0,x)}$.

  After presenting our results for the fluid velocity, the  energy density and the magnetic field as functions of both   space and time, we have validated our numerical calculation   by making a comparison with the experimental data\cite{phenix}. We see a great match between  the  hadron spectrum of our viscous calculation and the experimental results which is surprising because our model is very simple. We assume a conformal viscous fluid  which at early time behaves as a Guesser's fluid. Besides,  we have realized Guesser's solutions as an initial conditions for transverse fluid velocity and the energy density are quite reasonable.

  In our viscous calculations, we utilize two different methods to determine transport coefficients: Gauge/Gravity and pQCD. Interestingly, despite employing these distinct approaches, we achieve comparable outcomes, as depicted in Figs~\ref{f1} and~\ref{f2}.


\begin{thebibliography}{99}
\bibitem{na1} P. Romatschke, U. Romatschke, Viscosity information from relativistic nuclear collisions: how perfect is the fluid observed at RHIC? Phys. Rev. Lett. 99, 172301 (2007).
\bibitem{na2}  H. Song, U.W. Heinz, Causal viscous hydrodynamics in 2+1
dimensions for relativistic heavy-ion collisions. Phys. Rev. C 77, 064901 (2008).
\bibitem{na3}  P. Bozek, Flow and interferometry in 3+1 dimensional viscous hydrodynamics. Phys. Rev. C 85, 034901 (2012).
 \bibitem{na4}   C. Gale, S. Jeon, B. Schenke, P. Tribedy, R. Venugopalan, Event by-event anisotropic flow in heavy-ion collisions from combined Yang-Mills and viscous fluid dynamics. Phys. Rev. Lett. 110(1), 012302 (2013).
\bibitem{na5}   R. Andrade, F. Grassi, Y. Hama, T. Kodama, O. Socolowski Jr., On the necessity to include event-by-event fluctuations in experimental evaluation of elliptical flow. Phys. Rev. Lett. 97, 202302 (2006).
\bibitem{na6}   L. Del Zanna et al., Relativistic viscous hydrodynamics for heavy ion collisions with ECHO-QGP. Eur. Phys. J. C 73, 2524 (2013).	
\bibitem{na7} P. Kovtun, D. T. Son and A. O. Starinets, Phys. Rev. Lett. 94, 111601 (2005).
\bibitem{na8} J. Adams et al. [STAR Collaboration], Nucl. Phys. A 757, 102 (2005).

 \bibitem{a13}   W.T. Deng, X.G. Huang, Electric fields and chiral magnetic effect in Cu + Au collisions. Phys. Lett. B 	742, 296 (2015).
\bibitem{a14}  J. Adam et al., [ALICE Collaboration],Charge-dependent flow and the search for the chiral magnetic 	wave in Pb–Pb collisions at $\sqrt{s_{NN}} = 2.76$ TeV, Phys. Rev. C 93(4), 044903 (2016).
\bibitem{a15}  Q. Li et al., Observation of the chiral magnetic effect in ZrTe5. Nat. Phys. 12, 550 (2016).
\bibitem{a16}   D.E. Kharzeev, J. Liao, S.A. Voloshin, G. Wang, Chiral magnetic and vortical effects in high-energy nuclear collisions—a status report. Prog. Part. Nucl. Phys. 88, 1 (2016).
\bibitem{a17}  	 J. Xiong et al., Signature of the chiral anomaly in a Dirac semimetal: a current plume steered by a magnetic 	field,1503.08179.
	\bibitem{a18} M. H. Moghaddam, B. Azadegan, A. F. Kord, W. M. Alberico,  Non-relativistic approximate numerical
ideal-magneto-hydrodynamics of (1+1D) transverse flow in Bjorken scenario, Eur. Phys. J. C78, 255 (2018).
\bibitem{a19} M Haddadi Moghaddam, B Azadegan, AF Kord, WM Alberico,	Transverse expansion of hot magnetized Bjorken flow in heavy ion collisions, The European Physical Journal C 79, 1-12, (2019).
\bibitem{a20} AF Kord, A Ghaan, MH Moghaddam,    Analytical solutions of EM fields with the generalized Bjorken longitudinal expansion of QGP,  The European Physical Journal Plus 137, 1-20, (2022).
\bibitem{a21}  M.~Karimabadi, A.~F.~Kord, B.~Azadegan, Transverse expansion of Bjorken flow in (1+1D) relativistic ideal-magnetohydrodynamics with magnetization, Nuclear Physics A 1040, 122748 (2023).


\bibitem{a22}    V. Roy, A. K. Chaudhuri, and B. Mohanty, Comparison of results from a (2+1)-D relativistic viscous hydrodynamic model to elliptic and hexadecapole flow of charged hadrons measured in Au-Au collisions at $\sqrt{s_NN}=200 \ GeV$,  Phys. Rev. C86, 014902 (2012).
\bibitem{a23}  H. Niemi, G. S. Denicol, P. Huovinen, E. Molnar, and D. H. Rischke, Influence of a temperature-dependent shear viscosity on the azimuthal asymmetries of transverse momentum spectra in ultrarelativistic heavy-ion collisions,
Phys. Rev. C86, 014909 (2012).
\bibitem{a24}  L. Rezzolla, O. Zanotti, Relativistic Hydrodynamics, Oxford University Press, 2013.
\bibitem{a25} J. Goedbloed, R. Keppens, S. Poedts, Advanced magneto hydrodynamics  with applications to laboratory and astrophysical plasma
(Cambridge University Press)  2010.
\bibitem{a26} A.M. Anile, Relativistic fluids and magneto-fluids (Cambridge
University Press, Cambdridge) 1989.


\bibitem{new1} Paul Romatschke, and Ulrike Romatschke, Relativistic Fluid Dynamics In and Out of Equilibrium, 1712.05815v3 [nucl-th], 2019.
	
	
	
	
	
	
	
	\bibitem{new2} M. H. Moghaddam, B. Azadegan, A. F. Kord, W. M. Alberico,  Non-relativistic approximate numerical
ideal-magneto-hydrodynamics of (1+1D) transverse flow in
Bjorken scenario , Eur. Phys. J. C78, 255 (2018).
\bibitem{new3}  S.S. Gubser, Symmetry constraints on generalizations of Bjorken
flow. Phys. Rev. D 82, 085027 (2010).	

\bibitem{new4}    Alex Buchel. Resolving disagreement for eta/s in a CFT plasma at finite coupling. Nucl. Phys., B803:166-170, 2008, 0805.2683.
		
\bibitem{new5} Alex Buchel and Miguel Paulos. Relaxation time of a CFT plasma at 	finite coupling. Nucl. Phys., B805:59-71, 2008, 0806.0788.
 
\bibitem{new6}	 Alex Buchel and Miguel Paulos. Second order hydrodynamics of a CFT 	plasma from boost invariant expansion. Nucl. Phys., B810:40-65, 2009,  0808.1601.
\bibitem{new7}	 Omid Saremi and Kiyoumars A. Sohrabi. Causal three-point functions and 			nonlinear second-order hydrodynamic coefficients in AdS/CFT. JHEP, 		11:147, 2011, 1105.4870.
\bibitem{new8} Saso Grozdanov and Andrei O. Starinets. On the universal identity in second order hydrodynamics. JHEP, 03:007, 2015, 1412.5685.
\bibitem{new9} Saso Grozdanov and Andrei O. Starinets. Second-order transport, quasi-normal 	modes and zero-viscosity limit in the Gauss-Bonnet holographic	fluid. JHEP, 03:166, 2017, 1611.07053.

\bibitem{new10}    Guy D. Moore and Kiyoumars A. Sohrabi.  Thermodynamical  second order hydrodynamic coefficients. JHEP, 11:148, 2012, 1210.3340.
\bibitem{new11}     Peter Brockway Arnold, Caglar Dogan, and Guy D. Moore. The Bulk Viscosity of High-Temperature QCD. Phys. Rev., D74:085021, 2006, hep-ph/0608012.
\bibitem{new12}   Peter Brockway Arnold, Guy D Moore, and  and Laurence G. Yaffae.   Transport coefficients 
   in high temperature gauge theories. 2. Beyond leading log. JHEP,  05:051, 2003, hep-ph/0302165.
\bibitem{new13}    Mark Abraao York and Guy D. Moore. Second order hydrodynamic coefficients from kinetic theory. Phys. Rev., D79:054011, 2009, 0811.0729.
 \bibitem{new14}    Paul Romatschke and Dam Thanh Son. Spectral sum rules for the quark gluon
 plasma. Phys. Rev., D80:065021, 2009, 0903.3946.
 
 \bibitem{new14} B Sahoo, CR Singh, D Sahu, R Sahoo, and J Alam, Impact of vorticity and viscosity on the hydrodynamic evolution of hot QCD medium Bhagyarathi Sahoo, arXiv:2302.07668, 2023.

\bibitem{new15} Romuald A. Janik and Robert B.
 Peschanski. Asymptotic perfect 
fluid dynamics as a consequence of Ads/CFT. Phys. Rev., D73:045013, 2006, hep-th/0512162.
\bibitem{a27} S. Pu, V. Roy, L. Rezzolla and D. H. Rischke, Bjorken flow in one-dimensional
relativistic magnetohydrodynamics with magnetization , Phys. Rev. D 93, 074022 (2016).
\bibitem{cooper} F. Cooper and G. Frye, Phys. Rev. D 10, 186 (1974).
\bibitem{phenix}  K. Adcox et al., (PHENIX Collaboration), Formation of dense partonic matter in relativistic nucleus- nucleus collisions at RHIC:
Experimental evaluation by the PHENIX Collaboration. Nucl. Phys. A 757, 184 (2005).
\bibitem{Ratti}  C. Ratti, R. Bellwied, J. Noronha-Hostler, P. Parotto, I. Portillo Vazquez, J.M. Stafford. arXiv:1805.00088 [hep-ph].
\end{thebibliography}
\end{document}